\documentclass[a4paper,12pt]{article}
\pdfoutput=1
\usepackage{amsmath,amssymb,amsfonts}
\usepackage[dvips]{graphicx}
\usepackage{epsfig}
\usepackage{color}
\usepackage{verbatim}
\usepackage{calc}
\usepackage{bbm}
\usepackage{setspace}
\usepackage{array}
\usepackage{epstopdf}
\usepackage[caption=false]{subfig}
\usepackage[left=2.4cm,top=3.3cm,right=2.4cm,bottom=3.3cm,bindingoffset=0cm]{geometry}
\usepackage[colorlinks=true,citecolor=blue,urlcolor=blue]{hyperref}
\usepackage[square,numbers,compress]{natbib}

\newcommand{\beq}{\begin{eqnarray}}
\newcommand{\eeq}{\end{eqnarray}}
\newcommand{\bea}{\begin{eqnarray}}
\newcommand{\eea}{\end{eqnarray}}
\newcommand{\be}{\begin{equation}}
\newcommand{\ee}{\end{equation}}

\def\de{\partial}
\def\brc{\langle}
\def\ckt{\rangle}

\def\Tr{{\rm Tr}} 

\def\de{\partial}

\def\1{\mathbbm{1}}

\def\nn{\nonumber  \\}

\numberwithin{equation}{section}
\def\renormmu{M}
\def\phasemu{\xi}

\begin{document}

\title{
\begin{flushright}\ \vskip -1.5cm {\small {IFUP-TH-2019}}\end{flushright}
\vskip 20pt
\bf{ \Large Large-$N$ $\mathbb{CP}^{N-1}$ sigma model on a Euclidean torus:  uniqueness and stability of the  vacuum
}
}
\vskip 10pt  
\author{  Stefano Bolognesi$^{(1,2)}$, 
Sven Bjarke Gudnason$^{(3)}$,\\  Kenichi Konishi$^{(1,2)}$, Keisuke Ohashi$^{(3)}$    \\[15pt]
{\em \footnotesize
$^{(1)}$Department of Physics ``E. Fermi'', University of Pisa}\\[-5pt]
{\em \footnotesize
Largo Pontecorvo, 3, Ed. C, 56127 Pisa, Italy}\\[3pt]
{\em \footnotesize
$^{(2)}$INFN, Sezione di Pisa,    
Largo Pontecorvo, 3, Ed. C, 56127 Pisa, Italy}\\[3pt]
{\em \footnotesize
$^{(3)}$ Research and Education Center for Natural Sciences,  Keio University}  \\[-5pt]
{\em \footnotesize  Hiyoshi 4-1-1, Yokohama, Kanagawa 223-8521, Japan }
 \\ [5pt] 
    {\footnotesize stefanobolo@gmail.com,
      gudnason@keio.jp,
      kenichi.konishi@unipi.it,
      keisuke084@gmail.com}
}
\date{November 2019}

\maketitle

\begin{abstract}
In this paper we examine analytically the large-$N$ gap equation and its
solution for the $2D$ $\mathbb{CP}^{N-1}$ sigma model defined on a
Euclidean spacetime torus of arbitrary shape and size ($L, \beta)$, 
$\beta$ being the inverse temperature. 
We find that the system has a unique homogeneous phase, with
the $\mathbb{CP}^{N-1}$ fields  $n_i$  acquiring a dynamically
generated mass $\brc\lambda\ckt\ge\Lambda^2$ (analogous to the
mass gap of $SU(N)$ Yang-Mills theory in $4D$), for any
$\beta$ and $L$.  Several related topics in the recent literature are discussed. One concerns the possibility, which turns out to be excluded according to our analysis, of a ``Higgs-like'' - or deconfinement - phase at small $L$ and at zero temperature. Another topics involves ``soliton-like'' (inhomogeneous) solutions of the generalized gap equation, which we do not find.  A related question concerns a possible instability of the
standard $\mathbb{CP}^{N-1}$ vacuum  on ${\mathbb{R}}^{2}$, which is shown not to occur.  In all cases,  the difference in the conclusions can be traced to the  existence of certain zeromodes and their proper treatment.   The $\mathbb{CP}^{N-1}$ model with twisted boundary conditions is also analyzed.   The $\theta$ dependence and different limits involving
$N$, $\beta$ and $L$ are briefly discussed.
\end{abstract}

\newpage

\tableofcontents

\newpage

\section{Introduction}

The two dimensional $\mathbb{CP}^{N-1}$ sigma model has received
constant attention from theoretical physicists ever since the
pioneering works by D'Adda et.~al.~\cite{DDL} and Witten
\cite{Witten}.  See also  \cite{Eichenherr}-\cite{Bonanno:2018xtd}.\footnote{It is not our aim here to make a review of the vast literature on two dimensional $\mathbb{CP}^{N-1}$ sigma model; the references cited  in the text below are strictly relevant ones to the discussion. 
The list \cite{Eichenherr} - \cite{Bonanno:2018xtd} is certainly a partial and incomplete set of papers on the two dimensional $\mathbb{CP}^{N-1}$ sigma model.    }
The model is interesting as a toy model for nonperturbative dynamics
of QCD, sharing the properties of asymptotic freedom and confinement
with the latter. 
It can also be related to some physical phenomena in condensed matter
physics such as quantum Hall effects
\cite{Affleck:1984ar,Sondhi:1993zz,Ezawa:1999,Arovas:1999,Rajaraman:2002}.
Another context in which this model emerges as an effective action is
the internal quantum fluctuations of the nonAbelian vortex, in a
color-flavor locked $SU(N)_{\rm cf}$ symmetric vacuum
\cite{HananyTong,ABEKY,ShifmanYung,GSY,GJK}.

In spite of much effort dedicated to the study of the model, there
seem still to be some disagreement, new unconfirmed proposals,  and not
fully justified remarks in the current literature. It is our
purpose to address some of these issues
through a careful examination of the gap equations in a system
defined on a Euclidean torus of arbitrary size and form (finite spatial length $L$ and inverse-temperature $\beta$) in the large-$N$ expansion,
and try to resolve the controversies as much as possible. 

We find that the two dimensional $\mathbb{CP}^{N-1}$ sigma models defined with doubly periodic conditions,  possesses a unique
ground state for any  $L$ and $\beta$, which goes over, in the  $L\to \infty$ and at zero temperature ($\beta \to \infty$)  limit,  smoothly to the well-known vacuum with mass generation, and with no global $SU(N)$ symmetry breaking.  Our results turn out to differ from some claims found in the existent literature, and agree with some others. 
One concerns the possibility, which appears to be excluded according to our analysis, of a ``Higgs-like'' - or deconfinement - phase at small $L$ and at zero temperature. Another concerns ``soliton-like'' (inhomogeneous) solutions of the generalized gap equation, which we do not find.  As will be discussed in detail below (see in particular Section~\ref{tre}), in all cases the difference in the conclusions can be traced to the subtle role played by certain zeromodes 
(or in some cases, negative modes), and to their proper treatment.

%
%Even though our analysis below is a completely straightforward one, some of our results will turn out to differ from certain claims found in the existent literature. 
%The origin of such discrepancies will be carefully examined and eventually will be traced to the subtle role played by the zero or negative modes  in coupled nonlinear, saddle-point equations (gap equations) which determine the properties of the vacuum of our systems. 

The action of the ${\mathbb{CP}}^{N-1}$ sigma model is: 
\beq
S=    \int  dt dx \left[ \,   r\, (D_{\mu} n_i)^\dag (D^{\mu} n_i)  -  \lambda \, (n_{i}^\dag n_i - 1)      \, \right]   \,, \qquad  r\equiv \frac{4\pi}{g^2}\,,
\label{action}
\eeq
where $n_i$ with $i=1,\dots,N$ are 
%$N$
 complex scalar fields, and the
covariant derivative is
\be  D_\mu=\partial_\mu-i  A_\mu\,.    \ee
The action is invariant under the $U(1)$ gauge transformation
\be  n_i \to e^{i\alpha} n_i\,,  \qquad  A_{\mu} \to A_{\mu} -   \partial_{\mu} \alpha   \,.    \label{gaugetr}   \ee  
Classically  the $U(1)$ gauge field
$A_{\mu} $  can be integrated out, giving
\be     A_{\mu}  =  \frac{i}{2} \left( n_i^{\dagger} \de_{\mu}  n_i -   \de_{\mu} n_i^{\dagger}   n_i \right)\,, 
\label{acon}
\ee
 which upon insertion into the action  leads to the characteristic
 quartic interaction term among the fields $n_i$. 
$\lambda(x,t)$ is a Lagrange multiplier field enforcing the  classical
constraint
\beq
\label{classicalconstrain}
n_{i}^\dag n_i  = 1 \,.      \eeq
$ r\equiv \tfrac{4\pi}{g^2} $
is  the inverse of the coupling constant squared \footnote{Many
  different definitions of the coupling constant are used in the 
  literature on the $2D$ ${\mathbb {CP}}^{N-1}$ sigma model.  As this
  can potentially be confusing, we 
%review and
list   the relations among them in Appendix \ref{coupling}.}:  after an
appropriate rescaling of the $n_i$ fields  (see Eq.~\eqref{last}),  it  represents the
``size'' of the ${\mathbb{CP}}^{N-1}$ manifold.  

The ${\mathbb{CP}}^{N-1}$  model also admits the introduction of a $\theta$ term
\be    \Delta S =  \frac{i \theta }{2\pi}  \int d^2x \,  \epsilon_{\mu \nu}  \de_{\mu}  A_{\nu}  =   \frac{i \theta }{2\pi}  \int d^2x \,  F_{01} \equiv   -  i  \theta  Q\,,
\label{thetaterm}\ee
where $Q$ is the topological charge. $Q$ is classically quantized on ${\mathbb{R}}^{2}$ as 
\be   Q =  \frac{1}{2\pi i}  \int d^2 x  \, \de_{\mu}   \left(  n_i ^{\dagger}   \epsilon_{\mu \nu}    \de_{\nu} n_i \right) =   \frac{1}{2\pi}  \oint  dx_{\mu}  \, \frac{\de \sigma }{\de x_{\mu}}
\in {\mathbbm Z}\,,\label{topological}
\ee
where it is assumed that asymptotically $n_i =     e^{i \sigma} w_i $ with $w_i$ fixed and with $w_{i}^\dag w_i  = 1$ .

The paper is organized as follows.
In Section~\ref{sec:generaperiodic} we discuss the large-$N$ solution of the gap equation for the model defined on a Euclidean torus. In Section \ref{tre} we discuss two related issues  raised in the recent literature, regarding the uniqueness and  stability of the standard confining vacuum.
  In Section~\ref{sec:twist} the system with twisted boundary conditions is discussed.   In Section~\ref{theta} we make a brief remark on the dependence on the topological   $\theta$ angle.  We make concluding remarks in Section~\ref{discussion}.  Appendices~\ref{coupling}, \ref{sec:PauliVillars}   \ref{sec:formulas}, and  \ref{NSSB}   offer a collection of brief discussions on the coupling constant convention,    Pauli-Villars regularization, 
  various mathematical identities, and on the spontaneous symmetry breaking of the global $SU(N)$.

\section{\texorpdfstring{$2D$ ${\mathbb {CP}}^{N-1}$}{2D CP(N-1)}  sigma  model on  Euclidean 
%spacetime 
torus\label{sec:generaperiodic}}

\subsection{Analytic derivation of the gap equation \label{sec:analytic}}
Let us consider this model on a ring of size $L$ at a finite temperature $T= \beta^{-1}$.
The Euclidean partition function  $Z$  is
\begin{eqnarray}
Z&=&\int {\cal D}n \, {\cal D} \lambda \,  {\cal D} A \, \,  e^{-S_E - i \theta  Q }\;,\quad\nn
 S_E &=&\int_0^\beta dt \int_0^L dx \  \Big[ |D_t n_i |^2+ |D_x n_i|^2+\lambda(x) \left(|n_i|^2-r\right)
 +{\cal E}_{\rm uv}\Big]\;, \label{last}
\end{eqnarray}
where 
%\be  r= \frac{4\pi}{g^2}\,,
%\ee
%and 
bare energy density ${\cal E}_{\rm uv}$ is introduced.
Here fields $n_i(x,t)$ have the periodicity 
\begin{eqnarray}
n_i(x,t+\beta)=n_i(x, t)\,,\qquad n_i(x+L,t)=n_i(x, t)\,, \qquad  i=1,2,\ldots, N\, . \label{strict}
\end{eqnarray}
In this section we set $\theta=0$ (the $\theta$ dependence will be
discussed in Section~\ref{theta}).  
Assuming translational invariance,   $A_\mu(x,t)$ and $\lambda(x,t)$ can be
replaced by constants 
\begin{eqnarray}
  \langle A_\mu(x,t) \rangle  =a_\mu\,,    \quad
  \langle \lambda(x,t)\rangle= \lambda \,.
\end{eqnarray}
In the large-$N$ limit, one can omit contributions coming from their
fluctuations.   
The possibility that the $n_i$ fields acquire a nonvanishing classical VEV,
\be    \langle n_i(x,t)\rangle =\sigma_i\,,
\ee
will be discussed in Section~\ref{tre}.  For the moment, we set $\sigma_i = 0$. 

Integration over the constant  $\lambda$   (and $a_\mu$)
\begin{eqnarray}
Z=\int_{\mathbb{R}^2} d^2a_\mu \int^{i\infty +\epsilon}_{-i \infty +\epsilon}    d\lambda\,\ Z_\lambda\,,\qquad   {\rm def. } \,\  Z_\lambda\,,
\end{eqnarray}
can be  done  in the large-$N$ limit  by the saddle-point method:  
\beq
Z\simeq  Z_{\lambda}  |_{\lambda=\lambda_{\rm sp}},\qquad  \quad \frac{d \ln Z_\lambda}{d\lambda} \Big|_{\lambda=\lambda_{\rm sp}}=0\,.
\label{saddlep}
\eeq
Here in order to make the integral finite, the integration path for  $\lambda$ needs to be taken along the imaginary axis,  
whereas those for $a_\mu$ are along the real axis as usual. 
The pseudo free energy is
\beq
F_\lambda = -T \ln Z_\lambda \ .
\label{pfe}
\eeq
Sometimes the adjective ``pseudo''  will be used  to stress the fact that
$ F_\lambda$ it still a function of $\lambda$ and it is not yet extremized. When it acquires its stationary value 
\beq 
F = -T \ln Z\,,
\label{freal}
\eeq
 with $Z$ given in (\ref{saddlep}), one may refer to it as the  real free energy.  
When it is not specified  it should be clear from the context to which one we are referring to.

The strategy for solving the theory  in the large-$N$ limit is to first perform  the Gaussian integration over the $n_i$ fields,  and then to consider  the  saddle-point approximation (\ref{saddlep}).
After integrating out $n_i(x,t)$,  the partition function $Z_\lambda$   becomes
\begin{eqnarray}
-\ln Z_\lambda &=& N \sum_{n,m \in \mathbb Z}\sum_{I} c_I \ln \left( \left(\frac{2n \pi}{\beta }+a_t \right)^2+\left(\frac{2m \pi}L +a_x\right)^2+\lambda +\lambda_I \right)  \nn
&&\mathop+ \beta L ( -\lambda \, r+ {\cal E}_{\rm uv})\,, \phantom{\Big)} \label{eq:SMYpartition}
\end{eqnarray}
where $\lambda$ is determined by the saddle-point equation
\bea
0 &=& -\frac1{\beta L} \frac{\partial \ln Z_\lambda}{\partial \lambda} \nonumber \\
 &=& 
  \frac N {\beta L} \sum_{n,m \in \mathbb Z}\sum_{I} c_I \left( \left(\frac{2n \pi}{\beta } +a_t\right)^2+\left(\frac{2m \pi}L+a_x \right)^2+\lambda +\lambda_I \right)^{-1}  - r\,. \label{saddleeq}
\eea
Here we use Pauli-Villars regularization scheme
where  $c_I, \lambda_I$, with $I=0,1,2,3$, satisfy \footnote{
A simple possible choice  is
\begin{eqnarray}
c_1=1\,, \quad c_2=c_3=-1\,,\quad  \lambda_1= 2\Lambda_{\rm uv}^2\,,\quad \lambda_2=\lambda_3=\Lambda_{\rm uv}^2\;.
% \quad {\rm and~~} \Lambda_{\rm uv} \to \infty\,.
\label{eq:PVchoice}
\end{eqnarray}
} 
\begin{eqnarray}
& c_0=1\,, \quad \lambda_0=0\,,\quad  \lambda_{I\not=0} = b_I  \Lambda_{\rm uv}^2>0\,, & \phantom{\int_0)} \nonumber \\
& {\rm with} \ \ \ \sum_{I} c_I =0\,,\quad \sum_{I} c_I \lambda_I=0\,, &
\end{eqnarray}
which gives
\begin{eqnarray}
\sum_{I\not=0} c_I \ln \lambda_I= -\ln \Lambda_{\rm uv}^2+\sum_{I\not=0} c_I \ln b_I\,,\qquad  
\sum_{I\not=0} c_I \lambda_I \ln \lambda_I= \Lambda_{\rm uv}^2 \sum_{I\not=0} c_I b_I  \ln b_I\,.
\end{eqnarray}
$ \Lambda_{\rm uv}$ is the UV cutoff which  has to be sent  to infinity. See Appendix \ref{sec:PauliVillars} for more details.

In order to compute Eq.~\eqref{eq:SMYpartition} we first
use \footnote{The formula with $p\in \mathbb R_{>0}$  
\begin{eqnarray}
\frac1{p^s}=\frac1{\Gamma(s)}\int_0^\infty dt \;t^{s-1} e^{-t p}
\end{eqnarray}
is correct only for ${\rm Re}(s) >0$.   Using  Pauli-Villars
regularization and without any analytic continuation, 
 the above formula is extended with ${\rm Re}(s)>-2$ as is used in
 Eq.~\eqref{replace11}.  See Appendix \ref{sec:PauliVillars} for
 details.} 
\bea
\sum_I c_I \ln (\lambda'+\lambda_I) &=& \lim_{s\to 0} \sum_I c_I \frac{1-(\lambda'+\lambda_I)^{-s}} s  \nonumber \\
 &=&  -\lim_{s\to 0} \sum_I \frac{c_I}{s\, \Gamma(s)}  \int_0^\infty
 \,dt \, t^{s-1} \, e^{-t (\lambda'+\lambda_I)}\,, 
\eea
where $\Gamma(s)$ is the gamma function.
The key formula is the following identity 
\begin{eqnarray}
\sum_{n \in \mathbb Z } e^{-t \left(\frac{2\pi n}L+a\right)^2} =\frac{L}{2\sqrt{\pi t}} \sum_{n'\in \mathbb Z} e^{-\frac{\left(n' L\right)^2}{4t}+in' L a }\,.  \label{Identity}
\end{eqnarray}
which follows from the  Poisson
summation formula (see Appendix \ref{sec:formulas}).
By using these formulae, Eq.~\eqref{eq:SMYpartition} can be cast in
the form 
\bea
&& -\ln Z_\lambda - \beta L ( -\lambda r+ {\cal E}_{\rm uv})  \phantom{\int_0^\infty }\nonumber  \\
 &&   =-\lim_{s\to 0} \sum_{n,m,I} \frac{N c_I}{\Gamma(s+1)} \int_0^\infty  \,dt\; t^{s-1}\, e^{-t \left( 
 \left(\frac{2n \pi}{\beta }+a_t \right)^2+\left(\frac{2m \pi}L+a_x \right)^2+\lambda +\lambda_I \right) }\nonumber  \\
  & &  =-\lim_{s\to 0} \sum_{n',m',I} \frac{N \beta L\, c_I}{4\pi \Gamma(s+1)}  e^{i n' \beta a_t +im' L a_x}\int_0^\infty \, dt \; t^{s-2} \,e^{-  
 \frac{\left(n' \beta\right)^2+\left(m' L\right)^2}{4 t }- t (\lambda +\lambda_I) }\,.   \nonumber \\  \label{replace11}
\eea
We now note that, while the divergent part in the second line comes
from an infinite sum over $(n,m)$, the divergence in the last line
arises from the $n'=m'=0$ term alone, which thus can be separated
neatly from the rest. 
This term can be calculated and gives ($ \frac{N \beta L}{4\pi }$ times)
\begin{eqnarray}
&&\lim_{s\to 0}\sum_I \frac{c_I}{\Gamma(s+1)} \int_0^\infty   dt \, t^{s-2} \, e^{- t(\lambda+\lambda_I)}=
\lim_{s\to 0}\sum_I \frac{c_I \Gamma(s-1)}{\Gamma(s+1) (\lambda+\lambda_I)^{s-1}} \nn
&&\qquad =\lim_{s\to 0}\sum_I \frac{c_I (\lambda+\lambda_I)^{1-s}}{s(s-1)}
=\lim_{s\to 0}\sum_I  \frac{c_I(\lambda+\lambda_I)}{s(s-1)} \left( 1- s \ln (\lambda+\lambda_I) + {\cal O}(s^2) \right) \phantom{\int_0^\infty }\nn
&&\qquad = \sum_I c_I  (\lambda+\lambda_I) \ln  (\lambda+\lambda_I) \phantom{\int_0^\infty }\nn
&&\qquad = \lambda  \ln \frac{\lambda}e +  \lambda \sum_{I \not =0} c_I \ln \lambda_I + \sum_{I\not=0} c_I \lambda_I \ln \lambda_I 
+{\cal O}\left( \Lambda_{\rm uv}^{-2}\right)\,. \phantom{\int_0^\infty }
\end{eqnarray}
On the other hand, a generic $(n',m') \not =(0,0)$ term is equal  to ($ \frac{N \beta L}{4\pi }$ times)
\bea
&&   \int_0^\infty \, dt\; t^{-2} \,e^{- 
 \frac{\left(n' \beta\right)^2+\left(m' L\right)^2}{4 t } - t (\lambda +\lambda_I)  } \nn   && \qquad \quad  = \ 
 4 \sqrt{\frac{ \lambda+\lambda_I}{(n' \beta)^2+(m' L)^2}} \, K_1\left( \sqrt{ (\lambda +\lambda_I) ((n' \beta)^2+(m' L)^2)  } \right)\,,  \label{expression}
\eea
where $K_1(x)$ is the modified Bessel function of the second kind.    
Note that in the limit in which  the Pauli-Villars regulator masses  $\lambda_I$
($ I\ge 1$) are sent to infinity, this expression is nonvanishing only
for $I=0$. 
Therefore we find after relabeling $(n',m')\to (n,m)$,
\begin{eqnarray}
-\frac{\ln Z_\lambda}{\beta L} &=& -\frac{N}{4\pi}   \lambda \ln \frac{\lambda}{e \Lambda^2}+\renormmu\nn
&& -\sum_{(n,m) \in \mathbb Z^2 \setminus \{(0,0)\}} \frac{N}{\pi}
\cos(n \beta a_t ) \cos(m L a_x) \nn&&\qquad\times
\sqrt{\frac{ \lambda}{(n \beta)^2+(m L)^2}} \, K_1\left( \sqrt{ \lambda ((n \beta)^2+(m L)^2)  } \right),\qquad    \label{cosa}
\end{eqnarray}
where the renormalized constants $\Lambda, \renormmu$ are defined as
\begin{eqnarray}
r=\frac{N}{4\pi} \left( \ln \frac{\Lambda_{\rm uv}^2}{\Lambda^2}-\sum_{ I \not=0} c_I \ln b_I \right),\qquad  
{\cal E}_{\rm uv}=\renormmu+ \frac{N \Lambda_{\rm uv}^2}{4\pi}  \sum_{I \not=0} c_I b_I \ln b_I \,.
\label{re}
\end{eqnarray}
Being in an Euclidean space, this equation is invariant under the exchange of $(\beta, a_t)\leftrightarrow (L, a_x)$, as it should be. 
We note that the maximum  of  $Z_\lambda$  (i.e., the minimum of the free energy) with respect to $(a_t, a_x)$  for any given $\lambda$ is 
at   
\begin{eqnarray}
a_x =0\,,\quad a_t=0\,,
\label{mina}
\end{eqnarray}
as the coefficients of $\cos(n \beta a_t) \cos(m L a_x)$ in $\ln Z_\lambda$ are positive definite.
Thus we set $ a_x = a_t=0$, and find:
\begin{eqnarray}
-\frac{\ln Z_\lambda}{\beta L} &=& -\frac{N}{4\pi}   \lambda \ln \frac{\lambda}{e \Lambda^2}+\renormmu\nn
&&-  \sum_{(n,m) \in \mathbb Z^2 \setminus \{(0,0)\}} \frac{N}{\pi} 
 \sqrt{\frac{ \lambda}{(n \beta)^2+(m L)^2}} K_1\left( \sqrt{ \lambda ((n \beta)^2+(m L)^2)  } \right). \quad  \label{correspondsto}
\end{eqnarray}
The variation with respect to $\lambda$,  the gap equation  (\ref{saddleeq}),  becomes
%\be 
%0=-\frac1{\beta L N} \frac{\partial \ln Z_\lambda}{\partial \lambda } \,,      \label{gapeq0}\ee
%viz.,
\be   \frac{1}{4\pi} \ln \frac{\lambda}{\Lambda^2} =
\frac{1}{2\pi} \sum_{n,m \in \mathbb Z^2 \setminus \{0,0\}} K_0\left( \sqrt{ \lambda ((n \beta)^2+(m L)^2)  } \right). \label{gapeq}
\ee
This is an {\it exact} formula valid for any $(\beta, L)$, at
large-$N$.
For the model at the zero temperature, $\beta=\infty$, but with
generic value of $L$, only the $n=0$ term is present and 
Eq.~\eqref{gapeq} reduces exactly to the one found in
Ref.~\cite{Monin:2015xwa}.

\begin{figure}[h!t]
\begin{center}
\mbox{    
\includegraphics[width=3.2in]{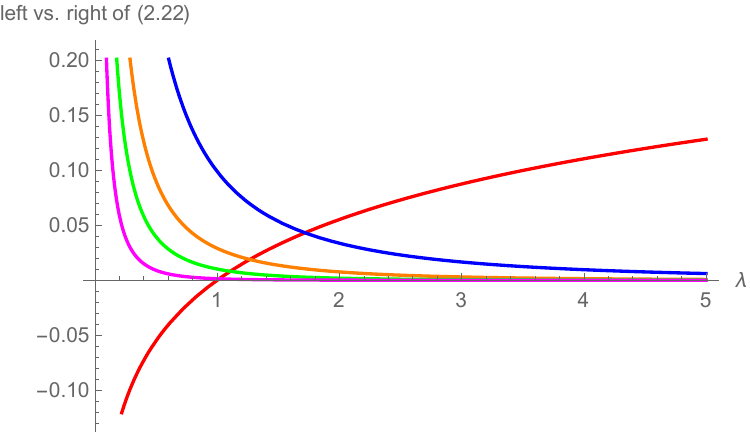}\quad
\includegraphics[width=3in]{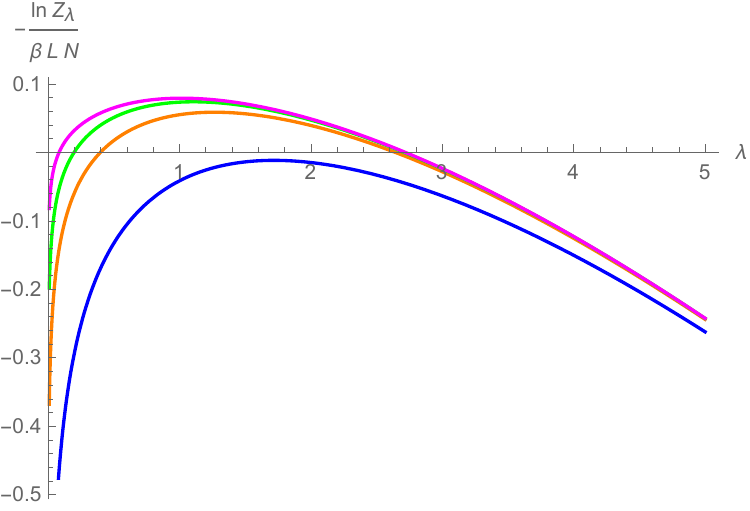}}  
\caption{\footnotesize Left panel: A graphic solution of
  Eq.~\eqref{gapeq} for $L=1$ and $\beta=1,2,3,5$, top to
  bottom, all measured in the unit of  $1/\Lambda^2$. The left-hand
  side of Eq.~\eqref{gapeq} is shown in red and the
  right-hand sides are shown in blue (top), orange, green,
  and magenta, respectively. 
  Right panel: The pseudo free energy normalized by $N$ times length
  $L$ as a function of $\lambda$.  $\beta=1,2,3,5$, from the bottom to
  the top.  The  real free energy is the maximum
  along the real axis of $\lambda$. 
} 
\label{gapeqFig}
\end{center}
\end{figure}
The left-hand side of Eq.~\eqref{gapeq}  is a monotonically increasing function of
$\lambda\in (0, \infty)$, varying from $-\infty$ to $\infty$, whereas
the right-hand side is a positive-definite function, monotonically
decreasing from $\infty$ to  $0$ (Fig.~\ref{gapeqFig}, the left panel).  
This equation therefore possesses a unique solution with nonvanishing
$\lambda$  such that 
\begin{eqnarray}
\lambda  \ge  \Lambda^2=\lim_{T\to 0, L\to \infty} \lambda\,,  
\label{minlam}
\end{eqnarray}
for any $(\beta, L)$ and  such that it attains its minimum value
$\Lambda^2$  in the case of the  standard $2D$  $\mathbb{CP}^{N-1}$  sigma model vacuum on $\mathbb{R}^2$ $(\beta=\infty, L=\infty)$.
The pseudo free energy is plotted in Fig.~\ref{gapeqFig}, the right panel. 
  This function is manifestly concave, as  $\frac{d^2\left(\ln Z_\lambda\right)}{d\lambda^2}>0$, which means that $F_{\lambda}$
  (\ref{pfe}), somewhat  counter-intuitively, is  {\it maximized}
  along the real axis of $\lambda$. See Subsection \ref{sec:Maximum} for a more general explanation of this fact

\begin{figure}[h!t]
\begin{center}
\mbox{    
\includegraphics[width=2.5in]{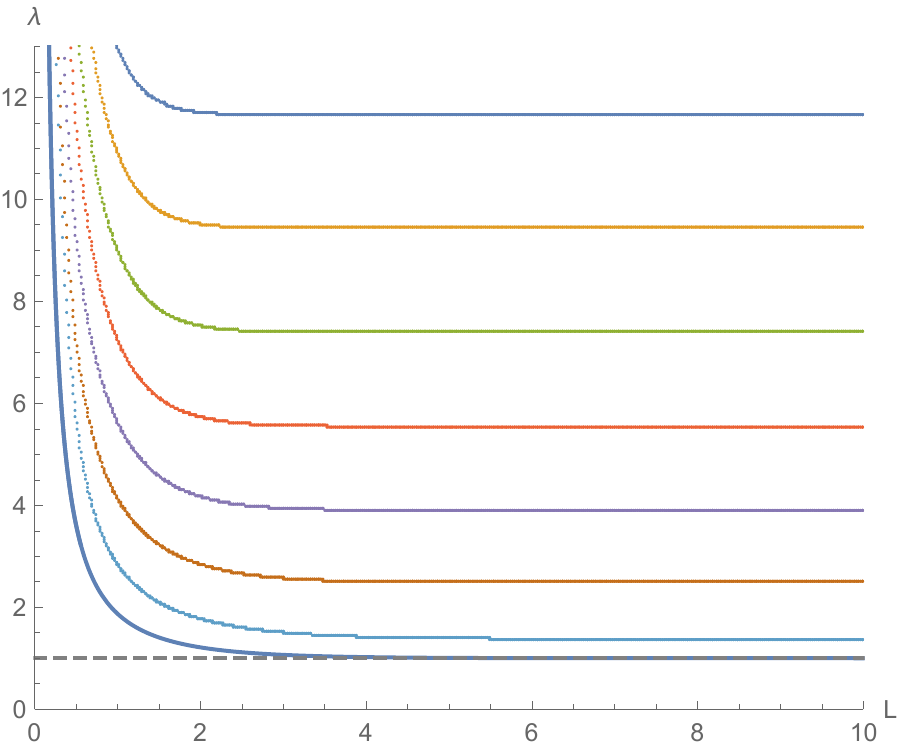}\qquad
\includegraphics[width=3.1in]{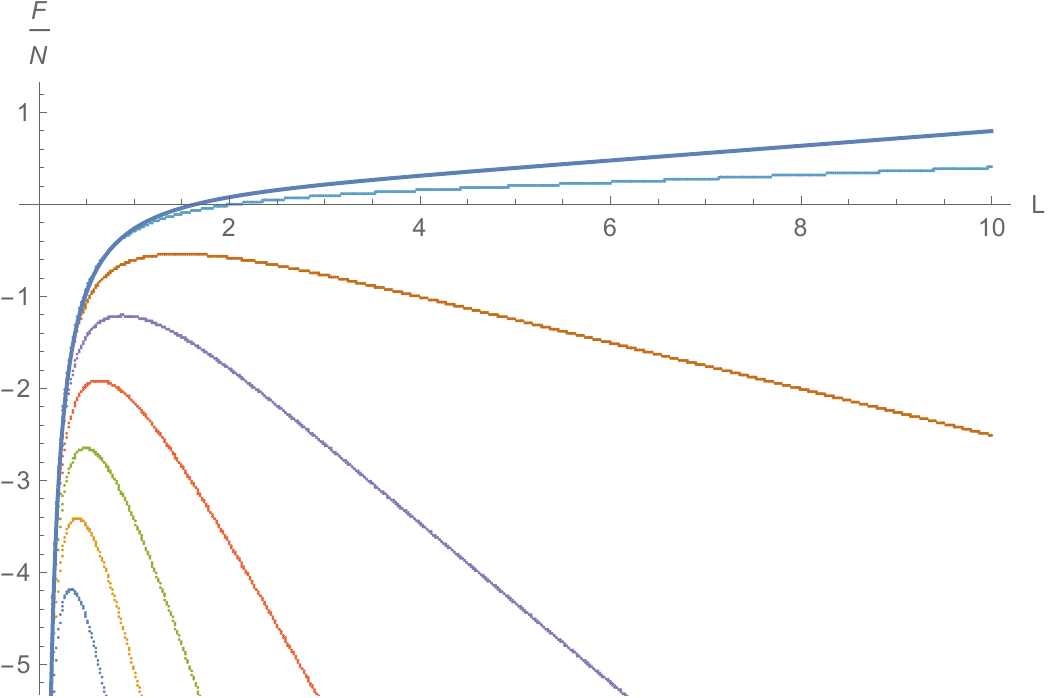}} 
\caption{\footnotesize Left panel: VEV of $\lambda$ as a function of $L$, with $\Lambda=1$ and $T=0, 0.5, 1,\dots, 3.5$, respectively, from the bottom to the top.  Right panel: the free energy  for  the same values and $\renormmu=0$ as functions of $L$.  In the right panel the curves refer to the temperatures
$T=0, 0.5, 1,\dots, 3.5$, from the top to the bottom.
} 
\label{figtwo}
\end{center}
\end{figure}

Configurations  of  $\lambda$ with various values of temperatures are shown in Fig.~\ref{figtwo}, the left panel.
Substituting the solutions into Eq. (\ref{correspondsto}) we obtain  the real free energy (\ref{freal}) shown in Fig.~\ref{figtwo}, the right panel.

%In the rest of the section we discuss two particular limits to illustrate some properties of the solution.

%%%%%%%%%%%%%%%%%%%%%%%%%%%%%%%%%%%%%%%%%%%%%%%%%%%%%%%%%%%%%%%%%%%%%

\subsection{Small size  and low temperatures}

%The above conclusion might sound somewhat counter-intuitive. 
%Let us %therefore
%study carefully the small size  limit of our solution. 
%Note that the gap equation requires  $\lambda > \Lambda^2$.
%If we take $\Lambda$ as a reference mass scale, we cannot obtain the
%deconfinement phase where $\lambda \to 0$.  
%Therefore, let us take $L^{-1}$ as the reference mass scale here,
We now study the  limit
\begin{eqnarray}
L \Lambda \to 0\,,\qquad  \beta \sqrt{\lambda} >\beta \Lambda  \to \infty\,,
\end{eqnarray}
(recall  $\lambda \ge  \Lambda^2$). 
Moreover we assume
\begin{eqnarray}
L \sqrt{\lambda} \to 0\,,
\end{eqnarray}
which will be verified a posteriori.
At low-temperatures  $T\ll \Lambda$,  the pseudo free
energy  is obtained from Eq.~\eqref{eq:SMYpartition} with $a_\mu=0$
as
\begin{eqnarray}
F_\lambda &\equiv &-T \ln Z_\lambda\nn
&= & N\sum_I c_I \sum_{n\in \mathbb Z} \sqrt{\left( \frac{2\pi n}L\right)^2+\lambda+\lambda_I}+L (-\lambda r+{\cal E}_{\rm uv})
+{\cal O}\left(e^{-\beta \sqrt{\lambda}}\right)\,.
\end{eqnarray}
 Using the expansion
\begin{eqnarray}
\sqrt{z+\lambda}=\sqrt{z}-\sum_{p=1}^\infty \frac{(-\lambda)^p}{z^{p-\frac12}}\frac{\Gamma\left(p-\frac12\right)}{2\sqrt{\pi} p!}\,,
\qquad  {\rm for} \ \  |\lambda|<z\,,
\end{eqnarray}
and the Riemann zeta function $\zeta(s)=\sum_{n=1}^\infty n^{-s}$, the
$n\not=0$ terms in 
$F_\lambda$ can be expanded for $\lambda$   small (compared to $\left (\frac{2\pi}{L}\right)^2$) as
\begin{eqnarray}
F_\lambda&=&-\frac{\pi N}{3L}+L \renormmu +N \sqrt{\lambda}-\frac{NL}{2\pi} \ln \frac{4\pi e^{-\gamma}}{\Lambda L}\, \lambda  \phantom{\int_0} \nn
&&-\frac{2\pi N }{L} \sum_{p=2}^\infty \frac{\Gamma\left(p-\frac12\right)}{\sqrt{\pi} p!}\left(-\frac{\lambda L^2}{4\pi^2} \right)^p \zeta(2p-1)
+{\cal O}\left(e^{-\beta \sqrt{\lambda}}\right)\,.
\end{eqnarray}
In the calculation of the constant and linear terms, the
regularization turns out to be crucial and we have used  
\begin{eqnarray}
\sum_I c_I \sum_{n\in \mathbb Z} \sqrt{\left(\frac{2\pi n}{L} \right)^2+\lambda_I}
&=&-\frac{\pi}{3L}-\frac{L}{4\pi} \sum_{I\not=0} c_I \lambda_I \ln \lambda_I\,,\nn
\sum_I c_I \sum_{n=1}^\infty \frac1{\sqrt{\left(\frac{2\pi n}{L} \right)^2+\lambda_I}}
&=&-\frac{L}{4\pi} \sum_{I\not=0} c_I
\left( \ln \frac{\lambda_I L^2}{16\pi^2}- 2\gamma \right)\;: \label{eq:InfiniteSums}
\end{eqnarray}
see Appendix \ref{sec:InfiniteSum} for more details. 

The gap equation for small $\sqrt{\lambda} L $ is given by
\begin{eqnarray}
0=\frac{\partial F_\lambda}{\partial \lambda}= \frac{N L}{4\pi}\left(\frac{2\pi}{\sqrt{\lambda}L}-2\ln \frac{4\pi e^{-\gamma}}{\Lambda L}
-\zeta(3)  \frac{\lambda L^2}{4\pi^2}+{\cal O}\left((\lambda L^2)^2\right) \right) \,,
\end{eqnarray} 
and its solution is
\begin{eqnarray}
\sqrt{\lambda}=\frac{2\pi}L \left(\frac1{2 \alpha}-\frac{\zeta(3)}{ 16 \alpha^4}+{\cal O}\left(\alpha^{-6}\right)\right),
\qquad  \alpha\equiv \ln \frac{4\pi e^{-\gamma}}{\Lambda L}\,,
\label{post}
\end{eqnarray} 
where $\alpha$ is a large parameter
%Here, taking $\sqrt{\lambda}L$ small and $\sqrt{\lambda} \beta$ large
%turns out to imply a small size $L\ll \Lambda^{-1}$ (large $\alpha$)
%and a low-temperature $ T\ll \Lambda$,
\begin{eqnarray}
1\, \gg\, \frac{\Lambda L}{4\pi e^{-\gamma}} =e^{-\alpha} \quad \big(\gg \,  e^{-\frac{\pi}{LT}}\big)\,.
\end{eqnarray}
Therefore, the free energy for small $L\Lambda$ is obtained as an  expansion in powers of $\frac{1}{\alpha} $:
\be   F=  -  T \ln Z=    -   \frac{\pi N}{3 L} \left(1-\frac3{2\alpha}% \frac1{\ln \frac{4\pi  e^{-\gamma}}{\Lambda L}}
+\frac{3\zeta(3)}{32 \alpha^4}
+{\cal O}\left(\alpha^{-6}, e^{-\frac{\pi}{LT \alpha}}\right) \right) +L \renormmu\,.
\ee
In the first term we recognize the L\"uscher term $  -   \frac{\pi N}{3 L} $,  with its thermal corrections, for the $2(N-1)\simeq 2N$ degrees of freedom which, due to asymptotic freedom, behave as if they were massless fields.  This agrees with the result obtained by Shifman et.al. \cite{Monin:2015xwa}  strictly at $T = 0$.  
%%%%%%%%%%%%%%%%%%%%%%%%%%%%%%%%%%%%%%%%%%%%%%%%%%%

\subsection{High temperature and large size}

\begin{figure}[!hbt]
\centering
\includegraphics[width=7.8cm]{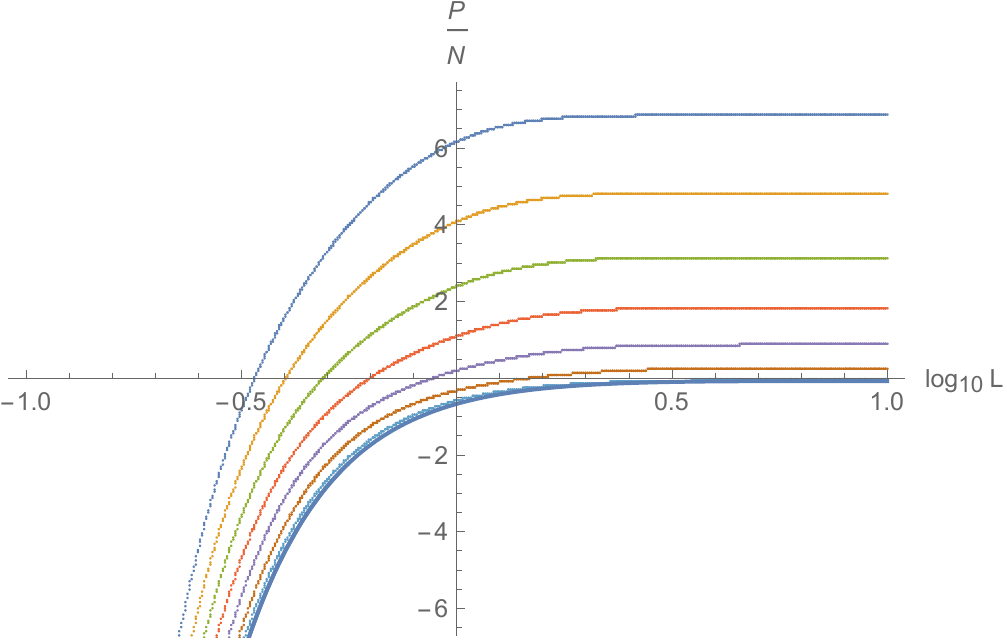} \quad 
\includegraphics[width=7.8cm]{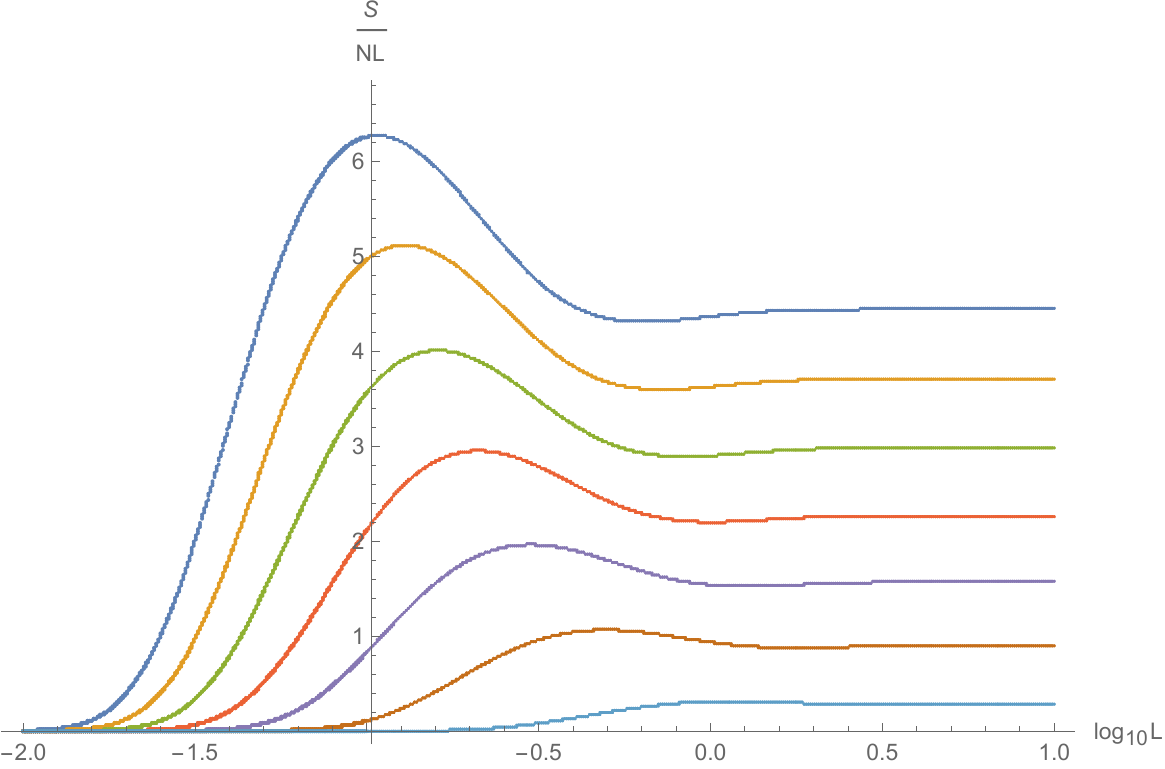}
\caption{Pressure and entropy density for $\Lambda=1, \renormmu=0$ and $T=0, 0.5, 1,\dots, 3.5$, from the bottom to the top.   } \label{fig:PS}
\end{figure}

The Euclidean torus possesses a symmetry
$L\leftrightarrow \beta =T^{-1}$. 
By using this we can relate  the case of high temperatures with large size with the one discussed in the previous subsection. 
Here we assume the limits
\begin{eqnarray}
\beta  \Lambda \to 0\,,\qquad  L  \sqrt{\lambda} >L \Lambda  \to \infty\,,
\end{eqnarray}
which implies also  $\beta  \sqrt{\lambda} \to 0$.
In these limits, a solution of the gap equation is obtained, from the dual of (\ref{post}), as
\begin{eqnarray}
\sqrt{\lambda}=2\pi T
\left(\frac1{2 \tilde \alpha}-\frac{\zeta(3)}{16 \tilde \alpha^4}+{\cal O}\left(\tilde \alpha^{-6}\right)\right),
\quad {\rm with} \ \ \tilde \alpha\equiv \ln \frac{4\pi e^{-\gamma} T}{\Lambda }\,,
\end{eqnarray}
and the free energy as
\begin{eqnarray}
   F=  -  T \ln Z=    -   \frac{\pi N}{3 } T^2L \left(1-\frac3{2\tilde \alpha}
+\frac{3\zeta(3)}{32 \tilde\alpha^4}
+{\cal O}\left(\tilde \alpha^{-6}, e^{-\frac{\pi LT}{\tilde  \alpha}}\right) \right) +L \renormmu\,.
\label{fapp}
\end{eqnarray}
The pressure and entropy of this system are then calculated as
\beq
 P\equiv -\frac{\partial F}{\partial L}\,, \qquad \quad  S\equiv -\frac{\partial F}{\partial T} \,,
\eeq
and are shown for various configurations in Fig.~\ref{fig:PS} for $F$ given in (\ref{fapp}).
In the extreme limit  $L\gg \Lambda^{-1}$,  $T\gg \Lambda$ we have
\bea
 P \simeq 
 \frac{\pi N}3 T^2-\renormmu\,, \qquad \quad  S  \simeq  \frac{2\pi N}3 T L \,.
\eea
This is to be compared with the other known limits
\begin{eqnarray}
 P\simeq 
\left\{
\begin{array}{c}
\displaystyle -\frac{ N \Lambda^2}{4\pi}-\renormmu \quad {\rm for~} L\gg \Lambda^{-1}, \quad T\ll \Lambda\,,  \phantom{\int_0^0}\\ 
\displaystyle -\frac{\pi N}{3L^2}-\renormmu  \quad {\rm for~} L\ll \Lambda^{-1},\quad  T\ll \Lambda\,,  \phantom{\int_0^0}
\end{array} \right.  \quad  S \simeq 0  \quad {\rm for~} T\ll \Lambda\,.
\end{eqnarray}

%%%%%%%%%%%%%%%%%%%%%%%%%%%%%%%%%%%%%%%%%%%%%%%%%%%%%%%%%%%%%%%%%%%%%%%%%%%%%%%%%%%%%%%%%%%%%%%%
%%%%%%%%%%%%%%%%%%%%%%%%%%%%%%%%%%%%%%%%%%%%%%%%%%%%%%%%%%%%%%%%%%%%%%%%%%%%%%%%%%%%%%%%%%%%%%%%%%%

\section{Uniqueness and stability of the
  \texorpdfstring{${\mathbb{CP}}^{N-1}$}{CP(N-1)} sigma model vacuum \label{tre}}

\subsection{Deconfinement (Higgs) phase at small \texorpdfstring{$L$}{L}? \label{question}}

It was argued in Ref.~\cite{Monin:2015xwa}   (see also \cite{HongKim})
that at small $L<L_{\rm crit} \sim \frac{1}{\Lambda} $   (at zero
temperature, $\beta= \infty$) the system undergoes a phase transition
into a deconfinement (or Higgs) phase \footnote{Even though they may not be entirely adequate, we stick for definiteness to the terminology used in \cite{Monin:2015xwa}, calling the standard  ${\mathbb{CP}}^{N-1}$  phase at zero temperature, $L=\infty$,  (\ref{standardCPN}), as "confinement" phase;  while using the word  "deconfinement (or Higgs) phase", for an eventual phase (\ref{without})   in which  no mass generation  for $n_i$ occurs, but  
with a non vanishing VEV for  ${\hat\sigma}^2  \equiv  \sum_i |n_i|^2$  signaling a spontaneous breaking of the global $SU(N)$ symmetry. 
 } where  
\be  \langle \lambda(x,t)  \rangle= \lambda =0\,, \qquad  \brc n_i \ckt =  \delta_{i 1}   \sigma \ne 0\,. \label{without}
\ee  
Consider the {\it  classical }equation of motion for the component $ \sigma$
\begin{eqnarray}
    \lambda\, \sigma=0\,, \label{eq:pathological}
\end{eqnarray}
that arises from the classical action \eqref{last}, with $\partial \sigma = 0$.
This appears to imply that two branches of solutions are possible, either $\lambda=0$ or $\sigma=0$. 
In the work \cite{Monin:2015xwa} they argued that the system has a solution \eqref{without} without
a mass gap, which becomes energetically favorable at small compactification length \footnote{This type of phase transition was first proposed in \cite{
Milekhin:2012ca} for the model on a finite interval with Dirichlet boundary condition. Such a possibility was  subsequently excluded \cite{BKO}.      
For models with mixed boundary conditions, however, see \cite{Milekhin2}, \cite{Pavshinkin}. 
},  as compared to  the standard   ${\mathbb{CP}}^{N-1}$ vacuum, 
\be    \brc  \lambda(x)  \ckt  =   \Lambda^2  \,,\qquad      \brc  \sigma  \ckt  =0\;, \qquad    (  L=\beta=\infty )\;. \label{standardCPN}
\ee

%A phase transition is, in principle, possible at large-$N$ but, for reasons we will explain below, is excluded in the present model.

%
%As a preliminary observation, note that an integration over the constant (in $x$ and $t$) modes $\sigma_i$
%gives 
%\begin{eqnarray}
%Z_\sigma\equiv \int \prod_{i=1}^N \, d\sigma_i \,d\bar \sigma_i \  e^{ - \beta L \lambda  |\sigma_i |^2}= \left( \frac{\pi}{\beta L \lambda}\right)^N\,.
%\label{eq:Zsigma}
%\end{eqnarray}
%This part of the partition function diverges as $\lambda \to 0$,  signaling that the vacuum with no dynamical mass generation for $n_i$, $\brc \lambda\ckt=0$, 
% cannot exist in the system. 
%and renders 
%a careful consideration of this limit indispensable.
%examination of the analysis done in \cite{Monin:2015xwa}. 

To examine whether a vacuum of the type,  (\ref{without}),   i.e.,  with a nonvanishing VEV for 
$ {\hat\sigma}^2  \equiv  \sum_i |n_i|^2  $ but without the dynamical mass generation,   $\brc \lambda \ckt =0$,      
can arise quantum mechanically at small $L$,   we rewrite the part of 
 the partition function due to the constant modes  as 
%arising from the constant modes: 
\begin{eqnarray}
Z_\sigma &=&\int d \hat \sigma^2    \int  \prod_{i=1}^Nd\sigma_i d\bar \sigma_i \   \delta\left(\sum_i |\sigma_i|^2-\hat \sigma^2\right)\,
 e^{ -\beta L \lambda \,\hat \sigma^2 }  \nn
&\simeq &  \int d\hat \sigma^2 \ \left(\frac{\pi e \hat \sigma^2}{N} \right)^N e^{-\beta L \lambda \hat \sigma^2}  = \int d\hat \sigma^2 \ 
e^{-S_{\hat \sigma}}\,,  \label{insert}
\end{eqnarray}
where we introduced an effective $\mathbb{CP}^{N-1}$ radius
$\hat\sigma\equiv \sqrt{ \sum_i |\sigma_i|^2}$,  and 
\begin{eqnarray}
S_{\hat \sigma} = N\ln \frac{N}{\pi e \hat \sigma^2}+\beta L \lambda \hat \sigma^2    \label{correctEffAct}
\end{eqnarray}
is the effective action for   $ {\hat\sigma}^2$.
The correct saddle-point equation for $\hat \sigma^2$ is therefore 
\begin{eqnarray}
 \hat \sigma^2 =\frac{N}{\beta L \lambda}\,,
\label{eq:xi_saddle}
\end{eqnarray}
which replaces   Eq.~(\ref{eq:pathological}).

%Inserted in Eq.~\eqref{insert} this saddle point reproduces the same
%result as Eq.~\eqref{eq:Zsigma}, 
%\begin{eqnarray}
%e^{-S_{\hat \sigma}}\Big|_{\rm saddle~pt.}=\left( \frac{\pi}{\beta L \lambda}\right)^N= Z_\sigma\;,
%\end{eqnarray}
%thanks to the large $N$ limit.  In this sense this saddle point is not
%pathological. 

To compare this saddle -point equation with
Eq.~\eqref{eq:pathological},  we rewrite Eq.~\eqref{eq:xi_saddle} as 
\begin{eqnarray}
      \lambda \,\hat \sigma = \frac {N}{\beta L \hat \sigma}\;.    \label{consistent}
\end{eqnarray}
The right-hand side comes from the volume integration of the
zero-modes.  Since there are $N$ copies of zero-modes, one cannot omit
this volume factor.
Therefore, for any finite $(\beta, L)$, there exists only one branch
of solutions of the coupled equations \eqref{eq:xi_saddle} and
\eqref{gapeq} and not two separate ones $\lambda=0$ or
$\sigma=0$.

One might object by saying that,  by first 
  taking the limit, e.g., $\beta \to \infty$,  Eq.~(\ref{consistent})
  yields indeed  $ \lambda \,\hat \sigma=0$.  
  As this point potentially involves a subtle question of ordering of various limits,  
 let us proceed with care.   
  Consider the
full partition function $Z_\lambda$:
\begin{eqnarray}
Z_\lambda =Z_{\rm mass} \, Z_\sigma =  \int d \hat \sigma \,  Z_{\rm mass} \,  e^{-S_{\hat \sigma}},
\end{eqnarray} 
where $Z_{\rm mass} $ is the part of the partition function found earlier by integration over massive modes (corresponding to the second term in 
Eq.~\eqref{correspondsto}).
The saddle-point equation for $\lambda$ is
\begin{eqnarray} 
0=-\frac{\partial \ln Z_{\rm mass}}{\partial \lambda} + \beta L \hat \sigma^2\,. \label{eq:deconf}
\end{eqnarray}
The first term in this expression does not contain contributions from the 
zero-modes. 
Note that at this point  the system of equations solved are  Eq.~(\ref{consistent}) and Eq.~(\ref{eq:deconf}),
valid in the large $N$ limit and at arbitrary  $(\beta, L)$.

Eq.~(\ref{eq:deconf})  is precisely  the form of the gap equation  (i.e., the saddle-point equation for $\lambda$) in the ``deconfinement'' phase,  
used in Ref.~\cite{Monin:2015xwa}.     
However,  the correct saddle-point equation for ${\hat \sigma}$ which accompanies it,  
 is  Eq.~\eqref{consistent},  and  not  Eq.~(\ref{eq:pathological}).  Use of   Eq.~\eqref{consistent}
 shows that  Eq.~(\ref{eq:deconf})   is equivalent to 
\begin{eqnarray}
0  &=&-\frac{\partial \ln Z_{\rm mass}}{\partial \lambda} + \frac{N}\lambda
=-\frac{\partial }{\partial \lambda} \left( \ln Z_{\rm mass}+N \ln \frac{\pi}{\beta L \lambda}\right)\nn
&=& -\frac{\partial \ln Z_\lambda}{\partial \lambda}\,,
\end{eqnarray}
which is exactly Eq.~\eqref{gapeq}, derived and studied in the previous section, and which has been shown to possess a unique nonvanishing
solution for $\lambda$, with $\lambda \ge \Lambda^2$,  for any values
of $(\beta, L)$.  
What Eq.~\eqref{eq:xi_saddle} tells us is that the two gap equations
in the ``deconfinement'' phase and in the ``confinement'' phase are
actually one and the same equation; its unique saddle point 
describes a single phase of the system, with dynamical generation 
of mass for the $n_i$ fields. A ``deconfinement'' phase with
$\brc\lambda\ckt=0$ never appears in our system \footnote{This
  conclusion is in agreement with the one reached in  Ref.~\cite{Munster, Ichinose},  but differs from that in Ref.~\cite{Monin:2015xwa}. 
  This latter fact can be traced to the proper treatment of the zeromodes, Eq.~(\ref{insert}), Eq.~(\ref{correctEffAct}). }.  
  The crucial point is that  the  coupled saddle-point equations for  $(\lambda, {\hat \sigma})$  have a smooth $\beta \to \infty$ limit.  
  
  Thus as far as we can see there is no problem of ordering in which different limits, $N, \beta, L \to \infty$, are taken, as long as 
   the vacuum property of the system at  $\theta=0$  is concerned.   As will be discussed in Section~\ref{theta},  the $\theta$ dependence 
  of the free energy, instead,  depends crucially on the order in which different limits are taken.

\begin{figure}[!ht]
\centering
\includegraphics[width=7.5cm]{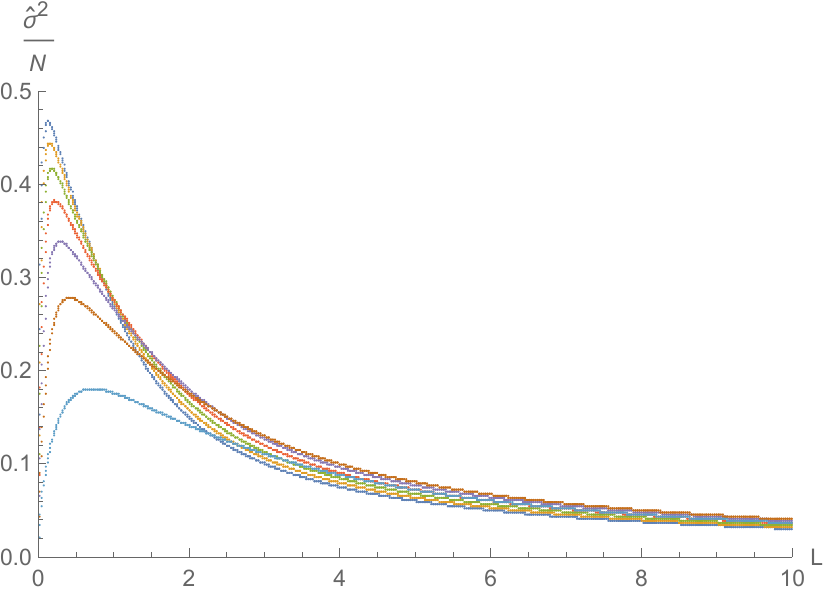} 
\caption{ \footnotesize VEV of $\frac{\hat \sigma^2}{N}$ for $\frac{T}{\Lambda}=0.5, 1, 1.5, \dots, 3.5$,  respectively, from the bottom to the top.  \label{fig:sigma} }
\end{figure}

The VEV  of  $\hat\sigma$ vanishes at zero
temperature due to the mass gap as 
\begin{eqnarray}
\hat \sigma^2  \le \frac{N T}{ L \Lambda^2}\,,\qquad  \quad  \lim_{T\to 0} \hat \sigma^2=0\,, \quad \forall L\;.
\end{eqnarray}
The volume integration of the large number of the zero-modes prevents
 $\hat\sigma^2$ from acquiring a nonvanishing VEV at zero temperature.    
Conversely, at any nonzero temperature, $T\not=0$, the mean square
 $\frac{\hat \sigma^2}{N}$ takes a nonvanishing expectation value 
(Fig.~\ref{fig:sigma}), but this is unrelated  to any
spontaneous symmetry breaking.   The absence of spontaneous breaking of the global $SU(N)$ symmetry in our system is discussed more explicitly in Appendix~\ref{NSSB}.

%%%%%%%%%%%%%%%%%%%%%%%%%%%%%%%%%%%%%%%%%%%%%%%%%%%%%%%%%%%%%%%%%%%%%%%%%%%%%%%%%%%%%%%%%%%%%%%%
\subsection{Absence of soliton-like solutions}

Another issue concerns the possible existence of inhomogeneous, soliton-like solutions of the gap equation.    It was recently argued \cite{GPV} that the standard
$2D$  ${\mathbb{CP}}^{N-1}$ sigma model vacuum,  
with dynamical mass generation
\be    \brc  \lambda(x)  \ckt  =   \Lambda^2  \,,\label{standard}
\ee
is actually unstable against decay into a lattice of solitons, i.e., inhomogeneous configurations $(n, \lambda)$  found in \cite{Nitta:2017uog}.  
These papers deal with the system at 
 zero temperature
($\beta=\infty$) and infinitely extended space ($L=\infty$).
The crucial issue, as in the previous Section \ref{question}, concerns a proper treatment of the zero-modes of the
$n_i$ fields.

In order to study a possible solution with a classical component,  $ n_i  =  \delta_{i N} \sigma$,  one first integrates the fluctuations of the quantum fields
$n_i$, $i=1, 2, \ldots, N-1$, to get an effective action
\be        S = (N-1)  \Tr  \log  \left(- \partial^2 + \lambda\right)  + \int d^2 x  \, \big[  (\partial \sigma)^2 + \lambda (\sigma^2 - r)   \big],  \label{similar}  
\ee
which, after functional variation with respect to $\lambda(x)$,  yields
\be      \sigma^2 =  r -   N   \, \left(  \int_{k \neq 0}   \frac{  |f_k(x)|^2}{ 2  \omega_k}   +   \sum_i \frac{  |f_{0 i}(x)|^2}{ 2  \omega_{0 i}}  \right)\,, \label{gap1}
\ee
\be     \left( - \partial^2 + \lambda(x) \right) f_k(x) = \omega_k^2 \,  f_k(x)\,.     
\ee
The last term in Eq.~\eqref{gap1} is the contribution from the eventual bound states $(f_{0 i}, \omega_{0 i})$:  it has been  separated from that of the continuum. 
Note that in any potential $\lambda(x)$  which asymptotically
approaches a value $\Lambda^2$ and  $\lambda(x)<\Lambda^2$ in some
region of $x$, there is at least one bound state of 
energy less than $\Lambda^2$, according to a known theorem in
one-dimensional quantum mechanics \cite{Schechter}. 
The variation with respect to the classical field $\sigma \equiv n_N$ gives 
\be  \left( - \partial^2 + \lambda(x) \right)  \,\sigma(x)=0\,. \label{gap2}
\ee   
The phase of the system corresponds to the solution of  the coupled equations (\ref{gap1}) - (\ref{gap2}).  These generalized gap equations have been discussed in detail in \cite{BKO} in the context of the ${\mathbb{CP}}^{N-1}$ defined on an interval with Dirichlet boundary conditions.
Note that problems may arise if the potential $\lambda(x)$ admits a bound state with zero energy, $\omega_0 = 0$: the equation (\ref{gap1}) would become ill defined due to an IR divergence.

The standard ground-state of the ${\mathbb {CP}}^{N-1}$  model  
corresponds to the homogeneous solution
\be    \sigma(x) \equiv 0 \,,   \qquad   \lambda(x)  \equiv \Lambda^2\,, \qquad     r  -   N   \,  \sum_{k}   \frac{  |f_k(x)|^2}{ 2  \omega_k}     =0\,.
\label{ordinarygap}\ee
A strict formula of the gap equation with Pauli-Villars
regularization is discussed in the next subsection. 
Note that in this case  the
spectrum  of  $n_i$   fields is purely a continuum.

In Ref.~\cite{Nitta:2017uog}, by using a map from the chiral Gross-Neveu model \cite{DunneBasar1,DunneBasar2} to the $\mathbb{CP}^{N-1}$ model, it is claimed that a soliton-like
configuration
\be  \lambda=   \Lambda^2 \left(  1 -  \frac{2}{\cosh^2 \Lambda x}  \right)\,, \qquad   \sigma(x)=   C  \, \frac{\Lambda}{\cosh \Lambda x} \,,\label{soliton}
\ee
where the center of the soliton $x_0$ is  taken at $0$,  satisfies  the
generalized gap equations (\ref{gap1}) - (\ref{gap2}). (More general solutions have  subsequently  been studied in \cite{NittaYoshii2}.)
 In Gorsky et.~al.~\cite{GPV} the energy density of this soliton solution was evaluated and found to be lower than the one of the standard homogeneous confining vacuum;  hence the claim of the potential instability of the standard vacuum.

Such a claim, however, is problematic, as  there is a  zero-energy bound state  for each $n_i$ field  in the potential (\ref{soliton}).  
The contribution from the latter in  Eq.~\eqref{gap1}  seems to be mysteriously missing  in their proof \cite{Nitta:2017uog,GPV}.

Note that in the ${\mathbb{CP}}^{N-1}$ system with Dirichlet boundary
conditions at $x=\pm \tfrac{L}{2}$ studied in
Refs.~\cite{BKO,BBGKO,BGKO},
the classical field $\sigma(x)$ solves  Eq.~(\ref{gap2}) and obeys the boundary condition
\be   
n \!\left(-\tfrac{L}{2}\right)=n\!\left(\tfrac{L}{2}\right) =
\sqrt{r}\,.\label{bordi}
\ee
In other words,   the ${\mathbb {CP}}^{N-1}$  constraint
\be     \sum_{i=1}^N     n_i^{\dagger} n_i  = r \,,
\ee
is saturated by the classical component $\sigma$ at the boundaries.
On the other hand, the quantum fluctuations are required to vanish at
the boundaries 
\be       n_i \!\left(\pm \tfrac{L}{2}\right)  =0\,,  \qquad i=1,2,\ldots, N-1\,.     \label{bdcond}
\ee
The difference in the boundary conditions \eqref{bordi} and
\eqref{bdcond},   explains why a zeromode $f_0(x)  \propto  \sigma(x)$
does not appear in the sum over quantum fluctuations of other
components $n_i$  ($i \ne N$) in the gap equation for the
finite-width  ${\mathbb{CP}}^{N-1}$  system \footnote{Of course,  it is crucial - and it was
  verified - that there are indeed no other zero energy solutions of
  $\left(-  \partial^2 + \lambda(x)\right)   f_0(x)=0$, satisfying the  boundary
  condition \eqref{bdcond}.}.   The classical function $\sigma(x)$,  although normalizable (it diverges logarithmically at the boundaries), does not belong to the domain of the self-adjoint operator.

In the background of the ``presumed'' soliton potential $\lambda(x)$ of
Eq.~\eqref{soliton}, the $n_i$ fields have one bound-state of
zero energy each,
\be    f_0(x) \propto  \frac{1}{\cosh  \Lambda
  x  }\,,\qquad  \omega_0=0\,, 
\label{zero-mode}
\ee
which is normalizable, satisfying the boundary condition
$f_0(x=\pm\infty)=0$, and therefore {\it must} be taken into account.
It is orthogonal to the continuum modes $f_k(x)$.  
The presence of the zero-mode contribution, the infrared divergent last
term in Eq.~\eqref{gap1}, means that the gap equation is not satisfied
by $\lambda(x)$ of Eq.~\eqref{soliton}.

Another way to show that  
Eq.~\eqref{soliton} does not constitute a solution,
is to consider a variational search for the solutions of the coupled
equations Eqs.~(\ref{gap1})-(\ref{gap2}), with respect to
$\lambda(x),\sigma(x)$.
For  $\lambda(x), \sigma(x)$ near the configuration
Eq.~\eqref{soliton}, there is a bound state $f_0(x)$ for each $n_i(x)$:
it is a near zero-energy bound-state mode for $n_i$:  there are no
reasons to omit  it from the sum in Eq.~\eqref{gap1}. As one approaches
the configuration Eq.~\eqref{soliton}, the failure of the gap equation 
Eq.~\eqref{gap1} is excerbated. The variational search for
$\lambda(x), \sigma(x)$ will drive one farther and farther away from
Eq.~\eqref{soliton}.

%We conclude that the standard  ${\mathbb{CP}}^{N-1}$ vacuum  (\ref{standard}) is stable. 
We conclude that the soliton  Eq.~(\ref{soliton}) is not a solution of the generalized gap equation \footnote{Another soliton-like configuration  was proposed in \cite{NittaYoshii2} in a twisted version of the ${\mathbb {CP}}^{N-1}$ model. 
 Their main result is in their Eq. (3.5), where the function $\lambda(x)$  is given by the known  P\"oschl-Teller potential,
 \be  \lambda(x)= -  \frac{2}{\cosh^2 x}\;. \label{PosTel} \ee 
This potential is actually the same as  Eq.~(\ref{soliton}) used in \cite{Nitta:2017uog},  just shifted by a negative constant. It means that  the zero-mode wave function Eq.~(\ref{zero-mode}) describes a {\it negative} mode in the potential Eq.~(\ref{PosTel}). 
It appears that the configurations considered there \cite{NittaYoshii2}  thus suffer from instability.}.    This also shows that the vacuum cannot be a lattice of solitons of this type, as suggested in \cite{GPV}. A more general reason for the uniqueness and stability  of the vacuum is given in the next two subsections. % \ref{sec33} and in Appendix \ref{sec:Maximum}.

%%%%%%%%%%%%%%%%%%%%%%%%%%%%%%%%%%%%%%%%
\subsection{Uniqueness of the saddle point on the real axis}
\label{sec33}

We now make an even stronger statement about the uniqueness of the solution of the gap equation.  We prove below that, under some generic assumptions, there are no solutions of Eqs.~(\ref{gap1})-(\ref{gap2})  other than the standard confining vacuum.

Let us consider an arbitrary real function 
$\lambda(x)$ (as well as
$A_x(x)$) with periodicity
$\lambda(x+L)=\lambda(x)$ and assume that all mass-squared eigenvalues of
\begin{eqnarray}
\left(-D_x^2+\lambda(x)\right) f_n(x)=\omega_n^2 f_n(x)\,,  \label{eq:eigeneq}
\end{eqnarray}
are positive definite: $\omega_n^2>0$.
We expect that the gauge fields vanish at the saddle point
and hence just set $A_t=0$, for simplicity.
After integrating the $n_i$ field fluctuations, the pseudo free energy  is
given by \footnote{We recall that $\lambda_I$'s denote the Pauli-Villars regulator masses (squared), as in Section~\ref{sec:generaperiodic}.}
\begin{eqnarray}
F_\lambda 
&=&T\sum_{(n,m)\in \mathbb Z^2}\sum_I c_I \ln\left( \left(2\pi n T\right)^2+\omega_m^2+\lambda_I \right) +\int_0^L dx \left( -\lambda(x) r+{\cal E}_{\rm uv}\right)\nn
 &=&\sum_{m \in \mathbb Z}  \sum_I c_I\, \rho\left(\omega_m^2+\lambda_I\right)   +\int_0^L dx \left( -\lambda(x) r+{\cal E}_{\rm uv}\right),
\end{eqnarray}
where the function $\rho(\lambda)$ is  defined as
\begin{eqnarray}
\rho(\lambda)=2T \ln \left(2\sinh\left(\frac{\sqrt{\lambda}}{2T}\right) \right),\qquad \lim_{T\to 0} \rho(\lambda)=\sqrt{\lambda}\,.
\end{eqnarray}
Here  the well-known formula \eqref{eq:harmonic} for a harmonic
oscillator with Pauli-Villars regularization has been used.
The gap equation for $\lambda(x)$ is given by
\begin{eqnarray}
0=\frac{\delta F_\lambda}{\delta \lambda(x)}\bigg|_{\rm saddle~pt.}=\sum_{n\in\mathbb Z} \sum_I c_I\rho'(\omega_n^2+\lambda_I)| f_n(x) |^2\bigg|_{\rm saddle~pt.} -r\;,   \label{eq:gengapeq}
\end{eqnarray}
where use was made of  the formula
\begin{eqnarray}
\frac{\delta \omega_n^2}{\delta \lambda(x)}=|f_n(x)|^2\,,\quad {\rm with~}\quad \int_0^L dx \bar f_n(x) f_m(x) =\delta_{nm}\,,
\end{eqnarray}
derived from Eq.~\eqref{eq:eigeneq}. Furthermore we have \footnote{ 
  For simplicity, here  the eigenvalues are  assumed to be non-degenerate,
  as $\omega_n^2\not=\omega_{-n}^2$ due to a non-trivial $A_x$.  
  The degenerate case is obtained by taking the limit
  $\omega_{-n}^2\to\omega_n^2$.  
  We also set diagonal components to zero, $u_n{}^n=0$, with an
  infinitesimal unitary transformation
  $\delta f_n(x)=i \sum_m u_n{}^m f_m(x)$  
  which is irrelevant for Eq.~\eqref{eq:dDelta}.} 
\begin{eqnarray}
 \frac{\delta f_n(x)}{\delta \lambda(y)}= \sum_{m(\not=n)} \frac{f_n(y)\bar f_m(y)}{\omega_n^2-\omega_m^2} f_m(x)\,.
\end{eqnarray}

Let us consider now a one-parameter family of functions interpolating  between two candidate gap functions $\xi_1(x)$ and
$\xi_2(x)$:
\be  \lambda(x)=(1-s)\,\xi_1(x)+s\,\xi_2(x) \;, \ee
and define
\begin{eqnarray}
\Delta_{nm} \equiv \int dx \frac{\partial \lambda(x)}{\partial s} f_n(x) \bar f_m(x)\,,
\end{eqnarray}
from which we find
\begin{eqnarray}
\frac{d \omega_n^2}{ds}=\Delta_{nn}\;,\qquad \frac{d \Delta_{nn}}{ds}= 2 \sum_{m(\not=n)}  \frac{|\Delta_{nm}|^2}{\omega_n^2-\omega_m^2}\,.  \label{eq:dDelta}
\end{eqnarray}
Now, the pseudo free energy $F_\lambda$ is % (also)
 a function of $s$,
and 
 we find 
\begin{eqnarray}
\frac{d^2F_\lambda}{ ds^2} =\frac{d}{ds} \sum_{I ,n}  c_I \rho'(\omega_n^2+\lambda_I) \Delta_{nn}= 
- \sum_I c_I \left( {\cal A}_I+ {\cal B}_I\right), \label{eq:Fss}
\end{eqnarray}
where  ${\cal A}_I$ and ${\cal B}_I$ are defined by 
\begin{eqnarray}
{\cal A}_I&\equiv& - \sum_{n\in\mathbb Z} \rho''(\omega^2_n+\lambda_I) |\Delta_{nn}|^2,\nn
 {\cal B}_I&\equiv & -\sum_{(n,m)\in \mathbb Z^2}^{n\not=m}
 \frac{\rho'(\omega_n^2+\lambda_I)-\rho'(\omega_m^2+\lambda_I)}{\omega_n^2-\omega_m^2} 
|\Delta_{nm}|^2\,.
\end{eqnarray}
We make now the following two natural assumptions:
\begin{eqnarray}
 {}^\exists C_\lambda, {}^\forall x:\,|\xi_1(x)-\xi_2(x)|  \le
 C_\lambda <\infty\,,\qquad
 \sum_n \frac1{\omega_n^2} <\infty\,.
\end{eqnarray}
It can then be shown that  the $I\neq 0$ terms vanish in the large
regulator-mass limit  
\begin{eqnarray}
\lim_{\Lambda_{I}\to \infty} {\cal A}_{I}=0,\quad\lim_{\lambda_{I}\to\infty}
    {\cal B}_{I}=0,\qquad {\rm for} \  I\neq 0\,,
\label{eq:ABvanish}
\end{eqnarray}
and ${\cal A}_0,{\cal B}_0$ have positive definite values (see Appendix \ref{Sec:ABvanish} for the proof).
Therefore, given two arbitrary different functions
$\xi_1(x),\xi_2(x)$, the pseudo free energy $F_\lambda$ is
a \emph{concave function} in $s$ \footnote{For constant $\lambda$ this result reduces to that found in Section~\ref{sec:analytic}, see Fig.~\ref{gapeqFig}.}:
\begin{eqnarray}
{}^\forall \xi_1(x), {}^\forall \xi_2(x), {}^\forall s\,: \quad  \lim_{\lambda_{I\not=0} \to \infty }
  \frac{d^2 F_\lambda}{ds^2}=-({\cal A}_0+{\cal B}_0) <0\,.  \label{absurdo} 
\end{eqnarray}

Let us now assume that $\xi_1$ and $\xi_2$ are two distinct solutions of the gap equation.  
In other words,  we assume,  for {\it reductio ad absurdum},  that the solution of the gap equation is not unique, and there are,  e.g.,  two solutions,  
$\xi_1$ and $\xi_2$.   The function $F$  then should satisfy
\be    \frac{d F_{\lambda}}{ds}\Big|_{s=0} =  \frac{d F_{\lambda}}{ds}\Big|_{s=1} =  0\;.
\ee
But such a function cannot exist, in view of Eq.~(\ref{absurdo}). 

We conclude that the solution of the gap equation \eqref{eq:gengapeq} is
unique and  therefore it must be the homogeneous vacuum
discussed in Section~\ref{sec:generaperiodic}.

\subsection{Maximization of the free energy \label{sec:Maximum}}

Formally  the pseudo free energy $F_\lambda$ must take  
{\it a local maximum} when evaluated on  a solution of the gap equation as a function of $\lambda$.
We checked this  explicitly for  formula \eqref{correspondsto} in Fig.~\ref{gapeqFig}. 
For an ordinary field, such as $n_i$ or $A_{\mu}$, this would signal the presence of a tachyonic instability
 but this is not the case for $\lambda$ which does not have a canonical kinetic term \footnote{Discussions on the quantum mechanically generated ``kinetic''  term for the field  $\lambda$ can be found in \cite{Novikov}.}.
% (not even dynamically generated \cite{}).
The integrand for the path integral can be regarded as a holomorphic function of $\lambda$.
Therefore, by fixing both the initial point ($\lambda=-i\infty+\epsilon$) and the final point ($\lambda=i\infty+\epsilon$),
the partition function is invariant under continuous transformations
of the path while avoiding any of poles.
Here we note that to make the partition function finite, 
the  path integration with respect to $\lambda$  must be along the imaginary axis for large $|\lambda|$.
A real configuration of $\lambda$ discussed here appears as an intersection between the real axis and such a path.
In this sense,  we can choose  any real VEV of $\lambda$  as long as the mass spectrum is positive definite.
(Here  one cannot  choose $\lambda$ so that  the mass spectrum contain zero-modes or negative modes 
         due to existence of poles as we noted).
When  we  apply the saddle-point approximation to this path integral, however,
     we have to use  a certain real VEV of $\lambda$  as a saddle point $\lambda=\lambda_{\rm sp}$  which satisfies the gap equation \eqref{saddleeq}.
(Here we assume that there is no {\it complex saddle point} which can contribute to the partition function.)
Around this saddle point,  it is natural to assume that $Z_\lambda$ behaves as
\begin{eqnarray}
\ln Z_\lambda=  {\rm const.} + a\, \beta  (\lambda-\lambda_{\rm sp})^2+\cdots\,,
\end{eqnarray} 
and the pseudo free energy  $F_\lambda$ must take the {\it minimum} at the saddle point along the imaginary axis,
\begin{eqnarray}
\frac{\partial^2 F_\lambda}{\partial ({\rm Im} \,\lambda)^2}\Big|_{\lambda=\lambda_{\rm sp}}=a  \quad 
\Rightarrow\quad a \in \mathbb R_{>0}\,.
\end{eqnarray}
Thus this feature means that,
formally,  the free energy must take a local maximum at the saddle point along {\it the real  axis},
\begin{eqnarray}
\frac{\partial^2 F_\lambda}{\partial ({\rm Re} \, \lambda)^2}\Big|_{\lambda=\lambda_{\rm sp}}=-a  <0\,.   
\end{eqnarray}
Therefore, counter-intuitively,  with a generic
$\lambda \in \mathbb R_{>0}$,
the pseudo free energy $F_\lambda$ can (or must) 
take  smaller  values than the real free energy  $F\equiv -T\ln Z$,
\begin{eqnarray}
F_\lambda  \le F_{\lambda_{\rm sp}} \simeq F\,.
\end{eqnarray}
A lesson which follows from these discussions is that even if one finds a field configuration with smaller free energy than that of  the vacuum  (such as the one in \cite{GPV},  or simply a generic configurations in Fig.~1, the right panel), this is not necessarily a signal of instability. 

%The same comment applies to the 
%

%%%%%%%%%%%%%%%%%%%%%%%%%%%%%%%%%%%%%%%%%%%%%%%%%%%%%%%%%%%%%%%%%%%%

\section {Twisted boundary conditions\label{sec:twist}}

In this section the analysis of Section~\ref{sec:generaperiodic} will be
repeated with a twisted boundary condition:
\be
     n_i(x,t+\beta)=   n_i(x, t)\,,\quad n_i(x+L,t)=  e^{ i \phasemu_i}  \,  n_i(x, t)\,, \qquad i=1,2,\ldots, N\,.  \label{twist}
\ee
By using the local $U(1)$ symmetry of the model, one can set
\be    \sum_i  \phasemu_i=0 \,.  
\ee
It is possible to define periodic fields, ${\tilde n}_i$, by
\be        n_i(x, t)  =   e^{i\frac{  \phasemu_i   x}{L} }  \,  {\tilde n}_i(x,t)\,,
\ee
where 
\be    {\tilde n}_i(x,t+\beta)=    {\tilde n}_i(x, t)\,,\quad  {\tilde n}_i(x+L,t)=  {\tilde n}_i(x, t)\,, 
\ee
but this introduces background gauge fields
\be  D_{\mu}  n_i = ( \partial_{\mu} +   i A_{\mu}  ) n_i =   e^{i\frac{   \phasemu_i   x}{L} }  \left(    \partial_{\mu} +   i A_{\mu}  + i  \delta_{\mu 2} 
 \frac{  \phasemu_i}{L}  \right)  {\tilde n}_i(x, t)\,, 
\ee
Dropping the tildes from now on, ${\tilde n}_i\to n_i$, we have
\be  S_E =\int_0^\beta dt \int dx \ \left( |D_t n_i|^2+ |D_x n_i|^2+\lambda(x) (|n_i|^2-r)\right)\,,
\ee
where 
\be  D_t   n_i   = \left(\partial_t +  i A_t \right) n_i \,, \qquad  D_x   n_i  =
\left( \partial_{\mu} +   i A_{x}  +   i  \frac{ \phasemu_i}{L} \right)   n_i\,.
\ee
Due to the periodicity of the system we can set the gauge fields
$A_t=  a_t$ and $A_{x} = a_x$ equal to constants, and also
set $\lambda(x,t) =\lambda$ constant. 
By integrating out $n_i(x,t)$, the partition function $Z_\lambda$ is given by
\begin{eqnarray}
-\ln Z_\lambda &=& \sum_i \sum_{n,m \in \mathbb Z}\sum_{I} c_I \ln \left( \left(\frac{2n \pi}{\beta }+a_t \right)^2+\left(\frac{2m \pi}L +a_x  +  \frac{\phasemu_i} L\right)^2   +\lambda +\lambda_I \right)  \nn
&&\mathop+ \beta L ( -\lambda r+ {\cal E}_{\rm uv})\,. \phantom{\Big)}  \label{eq:SMYpartitionTw}
\end{eqnarray}

For simplicity we consider only the ${\mathbb Z}_N$ symmetric form of   the phases $\phasemu_i$:
\be     \phasemu_i =   \frac{2\pi i}{N}\,, \qquad     i=1,2,\ldots, N\,. \label{twistzn}
\ee
A useful observation made in Refs.~\cite{DU,TinSule} is to   combine
the double sum     $\sum_i  \sum_m $ into a single sum   $\sum_k$ 
\be k=   N m + i \,, \qquad    k \in   {\mathbb Z} \,, \ee  
and the first line of Eq.~\eqref{eq:SMYpartitionTw} formally remains
the same, with the replacements
\be   L \to  N L\,,  \qquad    \sum_i  \to  1\,.
\ee  
We thus have
\begin{eqnarray}
-\ln Z_\lambda &=&     \sum_{n, k  \in \mathbb Z}\sum_{I} c_I \ln \left( \left(\frac{2n \pi}{\beta }+a_t \right)^2+\left(\frac{2 k  \pi}{N L} +a_x\right)^2+\lambda +\lambda_I \right)  \nn
&&\mathop+ \beta L ( -\lambda \,  r+ {\cal E}_{\rm uv})\,.  \phantom{\Big)}  \label{eq:SMYpartitionTw2}
\end{eqnarray}
The analysis of Section~\ref{sec:generaperiodic}  can now be repeated
step by step. Eq.~\eqref{replace11} is then replaced by
\bea
&& -\ln Z_\lambda - \beta L ( -\lambda r+ {\cal E}_{\rm uv})  \phantom{\int_0^\infty }\nonumber  \\
 && =-\lim_{s\to 0} \sum_{n,k,I} \frac{  c_I}{\Gamma(s+1)} \int_0^\infty  dt\;  t^{s-1} \, e^{-t \left( 
 \left(\frac{2n \pi}{\beta }+a_t \right)^2+\left(\frac{2 k \pi}{N L}+a_x \right)^2+\lambda +\lambda_I  \right) }\nonumber \\
  &&=-\lim_{s\to 0} \sum_{n',k',I} \frac{ \beta N L \, c_I}{4\pi \Gamma(s+1)}  e^{i n' \beta a_t +ik' N L a_x}\int_0^\infty  dt\;  t^{s-2} \, e^{- 
 \frac{\left(n' \beta\right)^2+\left(k' N L\right)^2}{4 t } - t (\lambda +\lambda_I)  }\,.      \nonumber \\ \label{replace1111}
\eea
Note that, remarkably, the explicit factor of $N$ which is absent in
the second line here -- as compared to the same line in
Eq.~\eqref{replace11} -- reappears in the third line  in front of the expression after
use of the crucial identity \eqref{Identity}. 
In the second line, divergences come from the infinite sum over $n$
and $k$, whereas in the last line, the divergent part arises from the
$n'=k'=0$ term only.  
The $n'=k'=0$ part (which is independent of the twisting) is given,
as before, by 
\be 
 \lambda \left(  \ln \frac{\lambda}e +\sum_{I \not =0} c_I \ln \lambda_I \right) + \sum_{i\not=0} c_I \lambda_I  \ln \lambda_I 
+{\cal O}( \Lambda_{\rm uv}^{-2})\,,
\ee
whereas  a generic  $(n',k') \not =(0,0)$ term is
\bea
&& \int_0^\infty  dt\; t^{-2} e^{- 
 \frac{\left(n' \beta\right)^2+\left(k' N L\right)^2}{4 t }- t (\lambda +\lambda_I) }\nn
 &&= 4 \sqrt{\frac{ \lambda+\lambda_I}{(n' \beta)^2+(k' N L)^2}} K_1\left( \sqrt{ (\lambda +\lambda_I) \left((n' \beta)^2+(k' N L)^2\right)  } \right)\,.
\eea
Note that only the $I=0$ term survives when the UV regulator masses
$\lambda_{1,2,3}$ are sent to $\infty$ ($\Lambda_{\rm uv}\to\infty$).
Therefore, we find, after renaming $(n',k')$ as $(n,k)$,
\begin{eqnarray}
-\frac{\ln Z_\lambda}{\beta L} &=& -\frac{N}{4\pi}   \lambda \ln \frac{\lambda}{e \Lambda^2}+\renormmu\nn
&& -\sum_{(n,k) \in \mathbb Z^2 \setminus \{(0,0)\}} \frac{N}{\pi} \cos(n \beta a_t ) \cos(k  N L a_x) \nn
&&\qquad \times  \sqrt{\frac{ \lambda}{(n \beta)^2+(k N L)^2}} K_1\left( \sqrt{ \lambda ((n \beta)^2+(k N  L)^2)  } \right),\qquad
\end{eqnarray}
where $\Lambda, \renormmu$ are determined by Eq.~(\ref{re}).
We again have the maximum $Z_\lambda$ with respect to $(a_t,a_x)$ with
a given $\lambda$ at $(0,0)$.
Thus we can set $a_x=a_t=0$ and 
\begin{eqnarray}
-\frac{\ln Z_\lambda}{\beta L} &=& -\frac{N}{4\pi}   \lambda \ln \frac{\lambda}{e \Lambda^2}+\renormmu\nn
&& -\sum_{(n,k) \in \mathbb Z^2 \setminus \{(0,0)\}} \frac{N}{\pi} 
 \sqrt{\frac{ \lambda}{(n \beta)^2+(k N  L)^2}} K_1\left( \sqrt{ \lambda ((n \beta)^2+(k N L)^2)  } \right). \quad\qquad
\end{eqnarray}
The gap equation is
\begin{eqnarray}
0&=&-\frac1{\beta L} \frac{\partial \ln Z_\lambda}{\partial \lambda }\nn
&=&-\frac{N}{4\pi} \ln \frac{\lambda}{\Lambda^2}
+\frac{N}{2\pi} \sum_{n,k  \in \mathbb Z \setminus \{(0,0)\}} K_0\left( \sqrt{ \lambda ((n \beta)^2+(k N  L)^2)  } \right).
\end{eqnarray}
This equation has a unique solution for $\lambda$,  with  
$ \lambda  \ge  \Lambda^2$,
as in the non-twisted system, (\ref{minlam}).

To conclude, the effect of the twist is to suppress the contributions
from $k>0$: the effect on the generation of the mass gap is
quantitative but not qualitative. 
The system with a small or finite $L$, behaves as one with a large
width, $N L\to\infty$.  This is consistent with the original idea of
the Eguchi-Kawai \cite{EguchiKawai} volume independence in $4D$ pure
YM theory, and in this context, we agree with \cite{TinSule}.

\section{\texorpdfstring{$\theta$}{Theta} dependence  of the free energy density \label{theta}}

The dependence of the energy density on the $\theta$ term
Eq.~\eqref{thetaterm} has been debated extensively in the literature.
We do not have much new to add here,  except for making a few
remarks for completeness.   

In the  standard extended $2D$ spacetime,  it was
explained  \cite{DDL,Witten,Affleck}  why  in the $\frac 1N$ expansion, the 
$\theta$ dependence appears at the leading perturbative order
$\mathcal{O}(\frac{1}{N})$ and is not exponentially suppressed, as
expected from the instantons,  by $e^{- c N}$.
To leading order in $\mathcal{O}(\frac{1}{N})$ the $\theta$ dependence of the
free energy density is  
\be      {\cal F} (\theta) =    \frac{3 \Lambda^2}{2 \pi}   \frac{\theta^2}{N} + \ldots\,,     \label{largeNtheta}
\ee
and this is consistent with the classical electric field energy,  
\be       \frac{E(\theta)}{L}  =   \frac{1}{2}    \left(   \frac  {e \theta}{2\pi}   \right)^2\,,
\ee
in the background electric field
\be      {\cal E}=    \frac{e \theta}{2\pi}
\ee
(or charges  $\pm \tfrac{\theta}{2\pi}$ at $x= \mp \infty$).
The coupling constant  
\be   e^2 =   \frac{12 \pi  \Lambda^2}{N} 
\ee
is inferred from the coefficient of the kinetic term $F_{\mu \nu}^2$ generated
by quantum effects (one-loop diagram of the $n_i$ fields).
%\cite{DDL,Witten,Affleck}.

Also, the nonanalyticity in $\theta$, a double vacuum degeneracy and a
spontaneous breaking of CP symmetry which occur at $\theta=\pi$, have been shown
to hold in the large-$N$ limit.
  Such a result has been confirmed
recently by using the ideas of generalized symmetry and mixed
anomalies \cite{GKKS}.

All this applies in the case of infinite line $L =\infty$ and at zero
temperature ($\beta=\infty$). As noted by many authors,  there
is a subtle issue of the different order in which the limits are
taken,  $\beta,  L \to \infty$, versus  $N\to \infty$.
Clearly, the well-known large-$N$ results reviewed above correspond to
taking the thermodynamic limit first, 
\be \lim_{N\to \infty}   \left( \lim_{\beta, L \to \infty} \, Z  \right)\,.   \label{largeN}
\ee

The $\theta$ dependence on an Euclidean torus with equal lengths ($L=\beta$) 
has been studied in Ref.~\cite{Asorey}, by summing over the topological
sectors. Their result can be easily generalized to our situation since it just depends on the  volume $V=\beta L$ and is thus 
 \be 
   {\cal F} (\theta) =  - \frac{1}{\beta L}      \log     \frac{ {\vartheta}_3\left(\frac{\theta}{2},  q\right) }
   {{\vartheta}_3(0,  q )}\,,  \label{formula} 
\ee
where
\be
  q\equiv  e^{i \pi \tau}\,, \qquad  \tau=   \frac{i N}{6 \beta L \Lambda^2 }\,,  \qquad  |q|<1\,,
\ee
and  $\vartheta_3(z, q)$  is Jacobi's theta function:
\be \vartheta_3(z, q) =   1 + 2 \sum_{n=1}^{\infty}  q^{n^2}   \cos  2 n z\,.
\ee
In the thermodynamic limit \eqref{largeN}, 
\be      N \ll   \beta L   \Lambda^2  \,, \qquad q = e^{- \frac{ \pi N} {6 \beta L \Lambda^2 } } \simeq 1\,,  \label{thermod}
\ee
one  first  uses the  re-summation formula
\be      \vartheta_3\left( z,    e^{ i  \pi \tau }\right)    =   \frac{\sqrt i}{\sqrt{\tau}} \,  e^{- \frac{ i z^2}{\pi \tau}}       \vartheta_3\left(\frac{z}{\tau},    e^{-\frac{i  \pi}{\tau}}\right)  \,,
\ee
and then takes the limit 
\be        e^{-\frac{i  \pi}{\tau}} =   e^{-  \frac{ 6  \pi  \beta L \Lambda^2 }{N}}  \to   0\,, \qquad  e^{- \frac{ i z^2}{ \pi \tau}}       \simeq   1 -    \frac{3\beta L \Lambda^2 }{2 \pi N}\,  \theta^2 +\ldots\,,
\ee
leading to the well-known large-$N$ result Eq.~\eqref{largeNtheta}.
Also a cusp at $\theta=\pi$ and a first-order transition between two
degenerate vacua is predicted \cite{Asorey}. 

In the opposite limit  in which 
\be      N \gg  \beta L   \Lambda^2  \,, \qquad q = e^{- \frac{ \pi N} {6 \beta L \Lambda^2 } }  \to 0 \,,  
\label{nfirst}
\ee
the formula \eqref{formula}   yields straightforwardly  an
instanton-like behavior,
\be         {\cal F} (\theta)  \simeq    \frac{1}{\beta L}     e^{- \frac{ \pi N }{  6 \beta L \Lambda^2}}   \,(1-  \cos \theta) + \ldots    \ . \label{instanton}
\ee
As pointed out in Ref.~\cite{Asorey},  most lattice studies of the
$\theta$ dependence in the $\mathbb{CP}^{N-1}$ model are done so far in the
regime of thermodynamical limit \eqref{thermod}, and  the instanton
behavior \eqref{instanton} which should appear in the limit (\ref{nfirst}), has not yet been corroborated, to the
best of our knowledge.

\section{Concluding remarks}
\label{discussion}

In this paper we examined the solution of the gap equation of
the large-$N$  $\mathbb{CP}^{N-1}$  sigma model on a Euclidean torus
of arbitrary shape and size,  $L$ and $\beta = T^{-1} $.  It was found that the system has  a unique 
ground-state for any $L, \beta$, in which a mass gap for the $n_i$ fields  
\be   \brc \lambda \ckt   \ge \Lambda^2 \,,
\ee
is generated. The system is in a phase analogous to the confinement
phase of QCD. In particular, a ``deconfinement'' phase proposed to
appear at small $L$ in some literature is shown not to occur in our
system. In particular  a spontaneous symmetry breaking never occurs. 
Another interesting proposal, that the system possesses
soliton-like inhomogeneous solutions of the gap equation, and that the
standard homogeneous ground-state of the $2D$ $\mathbb{CP}^{N-1}$  sigma
model is unstable against decay into a lattice of such soliton like solutions,  has been proven
not to be justified. 

We have presented a detailed discussion on the possible origin of the discrepancies found among  different literatures and with the present work. 
The principal cause for such differences is traced to the presence of certain zeromodes, whose proper treatment (according to us) leads to the said differences.

Concerning the ``deconfinement'' phase, a short comment on some
remarks made in Ref.~\cite{DU} is worthwhile.
In this paper, the phases  $e^{i\phasemu_i}$ appearing in the general
twisted boundary condition \eqref{twist}, are considered as further dynamical
degrees of freedom of the system.  Associated with the global symmetry
of the $\mathbb{CP}^{N-1}$ system,  the authors introduce the
concept of $SU(N)$ ``center symmetry'', which acts on the
``line operators'':
\be     e^{i  \oint  A_k}  \,, \qquad  A_k=   \frac{i}{2} \left(
n_k^{\dagger} \de_{\mu}  n_k-   \de_{\mu} n_k^{\dagger}  n_k \right)   \,
dx^{\mu}  \,, \quad (\textrm{no sum over }k %, \; k=1,2, \ldots N
)\,.
\ee
The periodicity conditions such as the strict one
\begin{eqnarray}
n_i(x,t+\beta)=n_i(x, t),\qquad n_i(x+L,t)=n_i(x, t)\;, \qquad  i=1,2,\ldots, N\;   \label{strictBis}
\end{eqnarray}
and a twisted one, 
\be
    n_i(x,t+\beta)=   n_i(x, t)\;,\quad n_i(x+L,t)=  e^{ i \mu_i}  \,  n_i(x, t)\;, 
\qquad     \mu_i =   \frac{2\pi i}{N}\;, \qquad     i=1,2,\ldots, N\;.      \label{twistBis}
\ee
considered in Section~\ref{sec:generaperiodic}  and Section~\ref{sec:twist}, respectively,  correspond to the VEV of $e^{i \mu_i}$ which
breaks the $SU(N)$ "center symmetry" (in the case of Eq.~(\ref{strictBis})), and {\it  vis a vis}, which is center symmetric  (for Eq.~(\ref{twistBis})).  
%Relying on an analogy with the physics of the $4D$ $SU(N)$ Yang-Mills
%theory, the authors of Ref.~\cite{DU} then state that for the strict periodic condition  (\ref{strict}) there is a ``deconfinement center broken phase'' at sufficiently small compactification length while this is not the case for the twisted boundary condition  (\ref{twist}),(\ref{twistzn}).
Relying on an analogy with the 
physics of the $4D$ $SU(N)$ Yang-Mills theory,   the authors of  \cite{DU} then state that the background  Eq.~(\ref{strictBis})  corresponds to 
a "deconfinement center broken phase", and the background  Eq.~(\ref{twistBis}) an analogue of "adjoint Higgs" (confinement?) phase. 
As shown in the main text here, however,  for  both boundary conditions the
dynamics of the system describes a confinement phase with dynamical
generation of the mass gap, $\brc\lambda\ckt \ge \Lambda^2$.  One has
the impression that,  in using the concepts such as center symmetry,
line operators, breaking of center symmetry as a criterion of a Higgs
phase, etc., the formal analogy with the  $4D$  YM theory has been
stretched too far, beyond what is justified by the actual analysis of
the dynamics. 
%\footnote{A relevant point which distinguishes two systems could be that while  the ${\mathbbm Z}_N$
%  ``center symmetry'' in the  $\mathbb{CP}^{N-1}$  model concerns only  a
%  global $SU(N)$ times local $U(1)$ gauge symmetry,   in the $4D$  $SU(N)$YM theory   the center ${\mathbbm Z}_N$  has to do with the strong  nonAbelian gauge dynamics of the system.
%}.
% 

In conclusion, the large $N$ $\mathbb{CP}^{N-1}$ model on a 2D torus  has a unique phase that  
looks like a confinement phase in the thermodynamic limit (large $L$ and zero temperature $\beta = \infty$). 
The mass gap $\lambda \ge \Lambda^2$ is generated for all $(\beta, L)$, and we find no phase transitions as 
$\beta$ and  $L$ are varied.  We critically commented on a number of papers in the literature, analyzing carefully the points which may 
have caused the discrepancies. In particular, soliton-like solutions  such as (\ref{standard}) \cite{Nitta:2017uog}, which are analogues of those discovered in the chiral Gross-Neveu model \cite{DunneBasar1,DunneBasar2} and which could have caused instabilities of the vacua \cite{GPV},  
 are actually found to be absent in the $\mathbb{CP}^{N-1}$ model.  
 Moreover, we have seen that a field configuration with smaller free energy than the vacuum, unless it is proven to be a
 correct solution of the gap equations,  is not necessarily a signal of instability if the field $\lambda$ is involved. 
 The  elegant map from the chiral GN model to the $\mathbb{CP}^{N-1}$ model
 used by Yoshii and Nitta \cite{Nitta:2017uog,NittaYoshii2,Nitta:2018lnn}, seems to fail subtly to produce the solutions of the gap equation for the latter from those
 in the former, due to the zero or negative modes which affect differently the two distinct physical systems.

%In the large-$N$ $\mathbb{CP}^{N-1}$ model compactifyed on a circle and at finite temperature ($2D$ Euclidean torus)   The system always remaind in a confined homogeneous vaccum with $ \brc \lambda \ckt  \neq 0$.
%The breaking of the $SU(N)$ symmetry by a could appear only by explicit breaking where a
%VEV of $n_i$ is fixed by some external factors.  
%The system which possesses a confinement phase and ones which admit a
%deconfinement phase (if they exist) are different systems.  No phase
%transition can connect them.

\section*{Acknowledgments}

We thank J.~Evslin, T.~Fujimori, M.~Nitta and R.~Yoshii  for useful discussions.
The work of S.B. and K.K.  is  supported by the INFN special 
project grant ``GAST (Gauge and String Theories)". 
S.B.G and K.O. are supported by the Ministry of Education, Culture, Sports, Science (MEXT)-Supported Program for the Strategic Research Foundation at Private Universities Topological Science (Grant No. S1511006).
S.B.G is also supported by Grant-in-Aid for Scientific Research on Innovative Areas “Topological Materials Science”
(KAKENHI Grant No. 15H05855) from MEXT, Japan.
K.O. is also supported by the Japan Society for the Promotion of Science (JSPS) Grant-in-Aid for Scientific Research (KAKENHI Grant No. 16H03984) from MEXT, Japan.
S.B.G. and K.O. thank  INFN and University of Pisa for the hospitality in November 2018 when part of this work was done.   K.K. and K.O. thank Prof. H. Itoyama for hospitality at the Osaka City University during their short stay there in December 2018, where part of the work was made.

\appendix

\section{The coupling constant}\label{coupling}

The action is
\be    S  =   \int d^2 x    \,  \left[  \,  {\cal C}   \,  (D_{\mu}  n_i)^\dag D_{\mu}  n_i\   - \lambda \,(  n_i^\dag  n_i   -1)
+  \frac{i \theta}{2\pi}  \epsilon_{\mu \nu}   \de_{\mu} A_{\nu}  \right],
\ee
where the  Lagrange multiplier $\lambda(x)$ is imposing the unit
radius of $\mathbb{CP}^{N-1}$.
One then proceeds to path integrate 
\be    W =     \int   {\cal D} n^*    {\cal D} n    \, {\cal D} A_{\mu}  \,{\cal D} \lambda   \,  e^{- S}    
\ee
to get the quantized version of the theory. 
The coefficient  ${\cal C}$  in front of the kinetic term for $n_i$
is the inverse bare coupling constant (or the square thereof);  
its choice is arbitrary and corresponds to a different definition of
the coupling constant adopted. 
Many different choices are used in the literature, some of which are
summarized in Table \ref{Babel}. 
\begin{table}[h]
  \centering 
  \begin{tabular}{|c|c|c|c|c|c|c|  }
\hline
% after \\ : \hline or \cline{col1-col2} \cline{col3-col4} ...
     Ref.       &  \cite{DDL}    & \cite{Witten,Affleck}    &  \cite{GSY,ABEKY,GJK}   &   \cite{Asorey}     &  \cite{DU}    &  \cite{TinSule}   \\
 \hline
  $\cal C$      &  $\frac{N}{2f}  $    &    $\frac{N}{g^2}   $   &   $\frac{4\pi}{g^2} $  &   $ \frac{N}{2 g^2} $  &   $\frac{2}{g^2}$    &  $\frac{N}{f}   $ \\[3pt]
\hline
\end{tabular}
  \caption{\footnotesize    Different definitions of the coupling constant used
    in the literature.  }\label{Babel} 
\end{table}
By rescaling the $n_i$ fields,  so as to render the kinetic terms of
$n_i$ canonical,  
\be    n_i \to      \frac{1}{\sqrt {\cal C} } n_i\,, \qquad   \lambda \to  {\cal C} \lambda \,, 
\ee
the action becomes
\be    S  =   \int d^2 x    \,  \left[  \,  \,  (D_{\mu}  n_i)^* D_{\mu}  n_i\   - \lambda \,(  n_i^*  n_i   -    {\cal C} )
+  \frac{i \theta}{2\pi}  \epsilon_{\mu \nu}   \de_{\mu} A_{\nu}  \right].  
\ee
In the large-$N$ approximation, after the functional integration over
the $n_i$ fields,  the gap equation shows the dynamically generated
mass of the $n_i$ fields which on  ${\mathbb{R}}^{2}$ is
\be      \sqrt{  \brc \lambda \ckt }  \equiv \Lambda = \mu \, e^{-\frac{   2\pi {\cal C}(\mu)}{ N}}  \,,    
\ee
which is $\mu$ independent.

There is no particular reason to prefer one  choice of ${\cal C}$ with respect to another.
The one we use here is inherited from the  nonAbelian vortex literature
%Gorsky et.~al.~\cite{GSY} and by us
\cite{ABEKY}-\cite{GJK}. Here the ${\mathbb {CP}}^{N-1}$ model emerges as a
long-distance  effective action describing the fluctuations of the
orientational zero-modes of the  nonAbelian vortex 
 \cite{HananyTong,ABEKY}, which emerges,  upon 
symmetry breaking of a $4D$ gauge theory,
\be    
\left(SU(N) \times U(1)\right)_{\rm c} \times SU(N)_{\rm f}  \,  {\stackrel  {v} {\longrightarrow}}   \, SU(N)_{\rm cf}        \label{Higgs}
\ee
in a color-flavor locked  Higgs  vacuum,  as a soliton excitation.  The factor  $\frac{4\pi}{g^2}$  in front of the ${\mathbb {CP}}^{N-1}$ action appears in a BPS saturated system, $g$ being the $4D$ gauge coupling constant at the 
mass scale  $v$  of the symmetry breaking \eqref{Higgs}.     From the
point of view of  the RG flow of the $2D$ ${\mathbb {CP}}^{N-1}$ model
at the mass scale  $\mu \le v$,    $g(v)$ plays the role of the UV
cut-off (bare) coupling constant.  Remarkably, the mass generated by
the $2D$  ${\mathbb {CP}}^{N-1}$  dynamics in the far infrared, 
\be    \sqrt{ \brc \lambda \ckt }  \equiv \Lambda = \mu \, e^{-   \frac{8  \pi^2}{ N g(\mu)^2}} \,,
\ee
coincides with the RG invariant mass of the $4D$   $SU(N)$  gauge
theory in the Coulomb phase ($v=0$).    In other words,  the $2D$
  dynamics of the $n_i$ fields appears to  ``carry on'' the RG flow of the $4D$
  $SU(N)$ gauge theory (which was frozen at $v$) below the mass
  scale $v$.
  This is a manifestation of the beautiful $2D$-$4D$  duality, first
  conjectured by N.~Dorey \cite{Dorey}  and subsequently confirmed
  explicitly in Refs.~\cite{HananyTong,GSY}.

%%%%%%%%%%%%%%%%%%%%%%%%%%%%%%%%%%%%%%%%%%%%%%%%%%%%%%%%%%%
\section{Pauli-Villars regularization}\label{sec:PauliVillars}
Let us consider the following Pauli-Villars regularization
\begin{eqnarray}
f(\lambda)\Big|_{{\rm reg}(\lambda)}\equiv  \sum_{I} c_I f(\lambda+\lambda_I)\,,
\end{eqnarray}
where  $\{c_i, \lambda_i\}$ is determined so that,   with a given   $p  \in \mathbb Z_{\ge 0}$,
\begin{eqnarray}
c_0=1\,, \lambda_0=0\,,\quad {\rm and~}\quad \lambda^k\Big|_{{\rm reg}(\lambda)}=\sum_{I} c_I \lambda_I^k=0\,,  \quad {\rm for~}  k=0,1,2,\cdots,p\,.
\end{eqnarray}
For instance,  with $p\ge 0$, we rediscover 
the well-known formula for the harmonic oscillator,
\begin{eqnarray}
&&\sum_{n\in \mathbb Z} \ln \left(
\left(\frac{2\pi n}{\beta} \right)^2+\omega^2 \right)  \Big|_{{\rm reg}(\omega^2 )} \nn
&& =\left. \ln \omega^2+ 2 \sum_{n=1}^\infty \left[ \ln \left\{1+ \left( \frac{\beta \omega}{ 2\pi n}\right)^2 \right\}
+ \ln \left(\frac{2\pi n}{\beta} \right)^2\right] \right|_{{\rm reg}(\omega^2 )} \nn
&&  =
\left. 2 \ln \left[ \omega \prod_{n=1}^\infty\left\{1+ \left( \frac{\beta \omega}{ 2\pi n}\right)^2 \right\}\right]\right|_{{\rm reg}(\omega^2 )} \nn
 &&  =\left. 2 \ln \left(2 \sinh \frac{\beta \omega}2 \right)  \right|_{{\rm reg}(\omega^2 )}. \label{eq:harmonic} 
\end{eqnarray}
Note that this identity is exact for arbitrary finite regulator masses.

%%%%%%%%%%%%%
\subsection{\texorpdfstring{$G_s(x,\epsilon)$}{Gs(x,epsilon)} and identities}
Let us define a function $G_s(x,\epsilon)$ as ($s  \in \mathbb C \setminus \mathbb Z_{\ge 0},  x,\epsilon\in \mathbb R_{>0}$),
\begin{eqnarray}
G_s(x,\epsilon)\equiv  \frac{1}{\Gamma(-s)}\int_\epsilon^\infty \frac{dt}{t^{s+1}}  e^{-x t}\,.
\end{eqnarray}
For ${\rm Re}\,s<0$, taking the limit of $\epsilon\to 0$ gives just a
familiar integration 
\begin{eqnarray}
\lim_{\epsilon\to 0} G_s(x,\epsilon)=x^s\,,\quad 
\Leftrightarrow\quad \frac1{\Gamma(u)}\int^\infty_0 dt t^{u-1} e^{-x t}=\frac 1{x^u}\quad {\rm for~} {\rm Re}(u)>0\,.
\end{eqnarray}
For ${\rm Re}\, s>0$, this limit gives divergences and the following
reduction is found 
\begin{eqnarray}
G_{s}(x,\epsilon)&=&\frac{1}{(-s)\Gamma(-s)} \left[\frac{e^{-xt }}{t^{s}} \right]^\infty_{\epsilon}+ 
 \frac{x}{(-s)\Gamma(-s)}\int_\epsilon^\infty \frac{dt}{t^{s}}  e^{-x t}\nn
 &=& x\, G_{s-1}(x,\epsilon)-\frac{1}{\Gamma(-(s-1))} \frac{e^{-x \epsilon}}{\epsilon^s}\,.
\end{eqnarray}
Therefore, we find, for ${\rm Re} s <0, k \in \mathbb Z_{> 0}$,
\begin{eqnarray}
G_{s+k}(x,\epsilon) =x^k G_{s}(x,\epsilon)- \sum_{n=1}^k \frac{x^{k-n}}{\Gamma(-(s+n-1))\epsilon^{s+n} }e^{-x\epsilon}\,,
\end{eqnarray}
and all divergent parts at $\epsilon =0$ are proportional to
\begin{eqnarray}
\frac1{\epsilon^{s+k}} (x \epsilon)^n\,, \quad  {\rm for~} n=0,1,2,\dots,k-1\,.  
\end{eqnarray}
Therefore, using Pauli-Villars regularization, the region of $s$ where
the limit of $\epsilon\to 0$ converges is extended as  
\begin{eqnarray}
&& \forall s \in \mathbb C \setminus \mathbb Z_{\ge0} |  {\rm Re}\, s < p+1 \phantom{\int}\nn
&&\int_0^\infty \frac{dt}{t^{s+1}}  e^{-\lambda t}\Big|_{{\rm reg}(\lambda)}
= \lim_{\epsilon\to 0} \Gamma(-s)G_s(\lambda, \epsilon)\Big|_{{\rm reg}(\lambda)} 
     = \Gamma(-s)\lambda^s \Big|_{{\rm reg}(\lambda)}\,.
\end{eqnarray}
Some integer cases,  $n=0,1,\cdots,p$, are defined  by taking the limit  as
\begin{eqnarray}
\int_0^\infty \frac{dt}{t^{n+1}}   e^{-\lambda t} \Big|_{{\rm reg}(\lambda)} &=&
\lim_{s\to 0} \lim_{\epsilon \to 0} \Gamma(-(n+s)) G_{n+s}(\lambda,\epsilon)\Big|_{{\rm reg}(\lambda)} \nn
&=&\lim_{s\to 0} \frac{\pi \lambda^{n+s}}{\sin(-\pi (n+s)) \Gamma(s+n+1)}  \Big|_{{\rm reg}(\lambda)}\nn
&=& \lim_{s\to 0} (-1)^{n+1} \frac{\lambda^{n+s}}{s\, n!}   \Big|_{{\rm reg}(\lambda)}
= (-1)^{n+1} \frac{\lambda^n}{n!}  \lim_{s\to 0} \frac{e^{s \log \lambda}-1} s \Big|_{{\rm reg}(\lambda)}\nn
&=&  \frac{ (-1)^{n+1}}{n!} \lambda^n  \log \lambda  \Big|_{{\rm reg}(\lambda)} .
\end{eqnarray}
Therefore, we find the useful identities
\begin{eqnarray}
\int_0^\infty \frac{dt}{t}   e^{-\lambda t} \Big|_{{\rm reg}(\lambda)} =& -\ln \lambda   \Big|_{{\rm reg}(\lambda)}
\quad &{\rm for~} p \ge 0\,,\nn
 \int_0^\infty \frac{dt}{t^\frac{3}2}   e^{-\lambda t} \Big|_{{\rm reg}(\lambda)}=& -2 \sqrt{\pi \lambda }  \Big|_{{\rm reg}(\lambda)}\quad&{\rm for~} p \ge 0\,,\nn
\int_0^\infty \frac{dt}{t^{2}}   e^{-\lambda t} \Big|_{{\rm reg}(\lambda)} 
=& \lambda \ln \lambda  \Big|_{{\rm reg}(\lambda)}\quad&{\rm for~} p \ge 1\,.
\end{eqnarray}
Here the right hand sides of these identities cannot be defined
without regularization. 

%%%%%%%%%%%%%%%%%%%%%%%%%%%%%%%%%%%%%%%%%%%%%%%%%%%%%%%%%%%%%%
\subsection{Eq.~\texorpdfstring{\eqref{eq:InfiniteSums}}{(2.30)}: Regularized infinite sums}\label{sec:InfiniteSum}
\def\Nuv{{N_{\rm uv}}}
Let us apply the Abel-Plana summation formula 
\begin{eqnarray}
\sum_{n=0}^{\Nuv} f(n)&=& \frac12 f(0)+ \frac12 f(\Nuv)+\int_0^\Nuv dx f(x)\nn
&&\quad +i \int_0^\infty dy  \frac{f(iy)-f(-iy)}{e^{2\pi y}-1} \nn
&&\quad -i \int_0^\infty dy  \frac{f(\Nuv+iy)-f(\Nuv-iy)}{e^{2\pi y}-1}\,,  \label{eq:AbelPlana}
\end{eqnarray}
to the function  $f(n)= \sqrt{n^2+\nu^2}$ by assuming $1,\nu \ll \Nuv$.
The first line of the right-hand side gives
\begin{eqnarray}
&&\frac{\nu}2 +\frac{\sqrt{\Nuv^2+\nu^2}}2+ \frac12 \left\{  \Nuv \sqrt{\Nuv^2+\nu^2}
+\nu^2 \ln \frac{\Nuv+\sqrt{\Nuv^2+\nu^2}}\nu \right\}\nn
&&= \frac{\Nuv(\Nuv+1)}2+\frac{\nu^2}2 \ln \frac{2\Nuv}\nu+\frac{\nu(\nu+2)}4+{\cal O}\left(\frac1{\Nuv^2}\right).
\end{eqnarray}
Due to the existence of branch cuts on the imaginary axis, the second line gives
\begin{eqnarray}
i \int_0^\infty dy  \frac{f(iy)-f(-iy)}{e^{2\pi y}-1}&=&-2 \int_\nu^\infty dy \frac{\sqrt{y^2-\nu^2}}{e^{2\pi y}-1}\equiv - {\cal K}(\nu)\,,
\end{eqnarray}
which vanishes in the large $\nu$ limit. 
Actually, using reparametrization of the integral variable
$y=\frac{\nu (t+t^{-1})}{2}$,  ${\cal K}(\nu)$ can be expressed in terms of 
the modified Bessel function of the second kind $K_1(x)$ as,
\begin{eqnarray}
{\cal K}(\nu)&=& \int_1^\infty \frac{dt}{2t}\left(t^2+\frac1{t^2}-2\right) \frac{\nu^2 }{e^{\pi \nu  (t+1/t)}-1}\nn
&=& \int_0^\infty \frac{dt}{2t}\left(\frac1{t^2}-1\right) \frac{\nu^2 }{e^{\pi \nu  (t+1/t)}-1}
= \sum_{n=1}^\infty  {\nu^2} \int_0^\infty \frac{dt}{2t}\left(\frac1{t^2}-1\right) e^{-\pi n \nu  (t+1/t)}\nn
&=& \sum_{n=1}^\infty \frac{\nu}{\pi n} K_1(2\pi n \nu)\,.
\end{eqnarray} 
The last term gives a constant term in  the large $\Nuv$ limit,
\begin{eqnarray}
&&2 \int_0^\infty dy \frac{{\rm Im} \sqrt{(\Nuv+iy)^2+\nu^2}}{e^{2\pi y}-1}
 \stackrel{ \Nuv\to \infty }=
2 \int_0^\infty dy \frac{y }{e^{2\pi y}-1}
=  {\cal K}(0)\,,\nn
&& {\cal K}(0)=2 \sum_{n=1}^\infty \int_0^\infty dy y e^{-2\pi n y}=\frac{1}{2\pi^2} \sum_{n=1}^\infty \frac1{n^2}=\frac1{12}\,,
\end{eqnarray}
where we used 
\begin{eqnarray}
{\rm Im} \sqrt{(\Nuv+iy)^2+\nu^2}=y \left(1+{\cal O}\left(\frac{\nu^2 }{\Nuv^2+y^2} \right) \right).
\end{eqnarray}
In summary, we get  the following result
\begin{eqnarray}
{\cal H}(\nu^2)&\equiv &\sum_{n=1}^\infty \left\{ \sqrt{n^2+\nu^2}-n-\frac{\nu^2}{2n} \right\}\nn
&=&\frac1{12}-\frac{\nu}2+\frac{\nu^2}2 \left(\ln \frac{2}\nu +\frac12 -\gamma\right)- \sum_{n=1}^\infty \frac{\nu}{\pi n} K_1(2\pi n \nu)\,,
\end{eqnarray}
where $\gamma$ is Euler's constant defined by
\begin{eqnarray}
\gamma \equiv \lim_{\Nuv \to \infty} \left( \sum_{n=1}^\Nuv \frac1n-\log \Nuv\right).
\end{eqnarray}
Using the above result,   the infinite sum $\sum_{n=1}^\infty n$ is regularized  ($p\ge 1$) as
\begin{eqnarray}
\sum_I c_I \sum_{n\in \mathbb Z} \sqrt{n^2+\lambda_I} &=& 
%\sum_I c_I \sqrt{\lambda_I}+2 \sum_I c_I \sum_{n=1}^\infty \left\{ \sqrt{n^2+\lambda_I}-n-\frac{\lambda_I}{2n}  \right\} \nn
\sum_I c_I \sqrt{\lambda_I}+2 \sum_I c_I {\cal H}(\lambda_I)\nn
&=&\sum_{I\not=0} c_I \left\{ \frac1{6}+\frac{\lambda_I}2\left(\ln \frac{4}{\lambda_I}+1-2\gamma \right)
+{\cal O} (e^{-2\pi \sqrt{\lambda_I} }) \right\}\nn
&\stackrel{\lambda_{I\not=0} \sim \infty}=&-\frac16 -\frac12 \sum_{I\not=0} c_I \lambda_I \ln \lambda_I\,,
\end{eqnarray}
where  $2\zeta(-1)=-\frac16$   appears as it should be. Similarly we find,
\begin{eqnarray}
\sum_I c_I \sum_{n=1}^\infty \frac1{\sqrt{n^2+\lambda_I}}&=&
\sum_I c_I \sum_{n=1}^\infty \left\{ \frac1{\sqrt{n^2+\lambda_I}}-\frac1n \right\}
=2\sum_{I\not=0}c_I \frac{d {\cal H}(\lambda_I)}{d \lambda_I}\nn
&=&\sum_{I\not=0} c_I \left\{ -\frac1{2\sqrt{\lambda_I}}
 +\frac12 \ln \frac4{\lambda_I}-\gamma +{\cal O}(e^{-2\pi \sqrt{\lambda_I}})\right\}\nn
 &\stackrel{\lambda_I\sim \infty}=& \gamma -\frac12 \sum_{I\not =0} c_I  \ln \frac{\lambda_I}4\,. 
\end{eqnarray}

%%%%%%%%%%%%%%%%%%%%%%%%%%%%%%%%%%%%%%%%%%%%%%%%%%%%%%%%%%
\subsection{Proof of Eq.~\texorpdfstring{\eqref{eq:ABvanish}}{(3.35)}}\label{Sec:ABvanish}
With the function 
\begin{eqnarray}
\rho(\lambda)=2\,T \ln \left(2\sinh\left(\frac{\sqrt{\lambda}}{2T}\right) \right) \qquad {\rm for~} \lambda >0\,,
\end{eqnarray}
one can show that function $\rho(\lambda)$  satisfies the following inequalities for arbitrary $\lambda >0$ and the limit
\begin{eqnarray}
\rho'(\lambda)>0\,,\quad \rho''(\lambda)<0\,,\quad \rho'''(\lambda)>0\,,  \quad (\lambda \rho''(\lambda))' >0\,, \quad \lim_{\lambda \to \infty} \lambda \rho''(\lambda)=0\,.
\end{eqnarray}
Especially, we find for $\lambda>0$, 
\begin{eqnarray}
-\frac{1}{\omega^2} \lambda \rho''(\lambda) >-\frac{1}{\omega^2+\lambda} \lambda \rho''(\lambda) \ge - \rho''(\omega^2+\lambda) >0\,.
\end{eqnarray}
Therefore, by assuming that $\sum_n \omega_n^{-2}$ is finite,  the
large regulator-mass limit gives 
\begin{equation}
\lim_{\lambda_I \to \infty}\sum_n \rho''(\omega_n^2+\lambda_I) =0\,.
\end{equation}
Using the inequality $\rho''<0$, we find 
\begin{eqnarray}
{\cal A}_I >0\;, \qquad  {\cal B}_I>0\,. 
\end{eqnarray}
Furthermore, with a natural assumption
\begin{eqnarray}
{}^\exists C_\lambda \in \mathbb R_{>0}\,, {}^\forall x\in [0,L] :\quad  |\partial_s \lambda(x)| \le C_\lambda\,,
\end{eqnarray}
we find $|\Delta_{nn}|$ has an upper bound as
\begin{eqnarray}
|\Delta_{nn}|=\left|\int dx \partial_s \lambda(x) |f_n(x)|^2 \right| 
\le \int dx |\partial_s \lambda(x)| |f_n(x)|^2  < \int dx C_\lambda |f_n(x)|^2 =C_\lambda\,,
\end{eqnarray}
from which an inequality for ${\cal A}_I$ is given
\begin{eqnarray}
0< {\cal A}_I=- \sum_{n\in\mathbb Z} \rho''(\omega^2_n+\lambda_I) |\Delta_{nn}|^2 < -C_\lambda^2
\sum_{n\in\mathbb Z} \rho''(\omega^2_n+\lambda_I)\,.
\end{eqnarray}
Thus we find the large regulator-mass limit of ${\cal A}_{I\not=0}$,  
\begin{eqnarray}
\lim_{\lambda_I \to \infty} {\cal A}_I  =0\,, \quad {\rm for~} I\not=0\,. 
\end{eqnarray}
The inequality $\rho'''(\lambda)>0$ gives,
\begin{eqnarray}
-\rho''(x)< -\frac{\rho'(x)-\rho'(y)}{x-y} <-\rho''(y)\,,\quad  {\rm for~} x>y\,,
\end{eqnarray}
and by using the completeness condition $\sum_n f_n(x)\bar f_n(y)=\delta(x-y)$, we find
\begin{eqnarray}
\sum_n  |\Delta_{nm}|^2 &=& \int dx dy \partial_s \lambda(x)  f_m(x)\left(\sum_n \bar f_n(x) f_n(y) \right)\bar f_m(y)  \partial_s \lambda(y)\nn
&=&\int dx  \left( \partial_s \lambda(x) \right)^2 | f_m(x)|^2 
< C_\lambda^2\,.
\end{eqnarray}
Combining these inequalities  we obtain a bound for ${\cal B}_I$ as  
\begin{eqnarray}
0< {\cal B}_I &=& -\sum_{(n,m)\in \mathbb Z^2}^{n\not=m}
 \frac{\rho'(\omega_n^2+\lambda_I)-\rho'(\omega_m^2+\lambda_I)}{\omega_n^2-\omega_m^2} 
|\Delta_{nm}|^2\nn
 &<&- 2\sum_{\omega_n^2>\omega_m^2}\rho''(\omega_m^2+\lambda_I) |\Delta_{nm}|^2 
 <- 2\sum_{m}\rho''(\omega_m^2+\lambda_I)\sum_n  |\Delta_{nm}|^2 \nn
&<&- 2\sum_{m}\rho''(\omega_m^2+\lambda_I)\times C_\lambda^2\,.
\end{eqnarray}
Finally, we also obtain the limit for ${\cal B}_{I\not=0}$ as
\begin{eqnarray}
\lim_{\lambda_{I}\to \infty} {\cal B}_I =0\,,\quad {\rm for~} I\not=0\,.
\end{eqnarray}

\section{Poisson summation formula}\label{sec:formulas}
Combining the identity
\begin{eqnarray}
\sum_{n\in \mathbb Z} e^{i n x} =2\pi \sum_{m\in \mathbb Z} \delta(x+2\pi m) \,,
\end{eqnarray}
and a Fourier  transformation
\begin{eqnarray}
   \tilde f(p)\equiv \int^\infty_{-\infty} \frac{dx}{\sqrt{2\pi}}   f(x) e^{-ipx}\,,
\end{eqnarray}
the Poisson summation formula is proved:
\begin{eqnarray}
\sum_{n\in \mathbb Z} \tilde f(n q)e^{i n \theta }&=&\sum_n \int^\infty_{-\infty} \frac{dx}{\sqrt{2\pi} } f(x) e^{-i(x q-\theta ) n}\nn
&=&\frac1{|q|}\sum_n \int^\infty_{-\infty} \frac{dy}{\sqrt{2\pi} } f\left(\frac{y+\theta }q\right) e^{-i y n}\nn
&=&\frac{\sqrt{2\pi}}{|q|} \sum_{n\in \mathbb Z} f\left(\frac{2\pi n+\theta }q\right).
\end{eqnarray}
Applying  the Poisson summation formula to a  Fourier transformation of the Gaussian function $f(x) =e^{ -t (x/L)^2}$ 
 \begin{eqnarray}
\tilde f(p)=\int^\infty_{-\infty} \frac{dx}{\sqrt{2\pi}} e^{- t \left(\frac{x}L\right)^2-i p x }=\frac L {\sqrt{2 t}} e^{-\frac{(p L)^2}{4 t}}\,, \quad {\rm for~} L, t \in \mathbb R_{>0}\,,
\end{eqnarray}
 we  obtain 
\begin{eqnarray}
\sum_{n \in \mathbb Z } e^{-t \left(\frac{2\pi n}L+a \right)^2} =\frac{L}{2\sqrt{\pi t}} \sum_{n\in \mathbb Z} e^{-\frac{(n L)^2}{4t}+in L a }\,.
\end{eqnarray}
For instance, 
 the property of the theta function $\theta(x)$ is shown as,
\begin{eqnarray}
 \theta(x) = \sum_{n \in \mathbb Z} e^{-n^2 \pi  x} =\frac1{\sqrt{x}} \sum_{n \in\mathbb Z} e^{- \frac{n^2\pi} x} = \frac1{\sqrt{x}} \theta(x^{-1})\,,
\end{eqnarray}
with $x>0$.

\section{Absence of spontaneous breaking of the global $SU(N)$ symmetry \label{NSSB}}

%The standard way to see if spontaneous symmetry breaking occurs or not is as  follows.

Let us add the source term for $n_i$:
\begin{eqnarray}
\Delta S_J= -\int d^2x  ( \vec n \cdot \vec J^\dagger +{\rm h.c} )\;, 
\end{eqnarray}
then 
\begin{eqnarray}
\langle \vec n(x) \rangle_J  \equiv   \frac{\delta }{\delta \vec J(x)^\dagger} \ln Z[J]
\;;\qquad  
   Z[J] = \int D A_\mu \,D \lambda  \, D n\, D n^\dagger \, e^{-S-\Delta S_J}.
\end{eqnarray}
Path integration over the $\vec n$ fields includes the constant modes $\vec \sigma$.  
The spontaneous symmetry breaking occurs if
\begin{eqnarray}
\lim_{J\to 0}  \langle \vec n(x) \rangle_J \ne 0\,. 
\end{eqnarray}

Let us take $J(x)$ to be constant $\vec J(x)=\vec J_c$,    which  is sufficient for our purpose.
Also the translational-invariance assumption is made. 
Then one can easily calculate and obtain the partition function and  the (pseudo-) free energy as
 \begin{eqnarray}
 \ln Z[J] \sim \ln Z_\lambda[J]=\ln Z_\lambda +\frac{|\vec J_c|^2}{\lambda} L \beta\\
F[J]\sim F_\lambda[J]=F_\lambda  -\frac{|\vec J_c|^2}{\lambda} L
\end{eqnarray}
with $Z_\lambda, F_\lambda$ given in  Eq.(2.20),   Eq.(2.23)  or  Eq.(2.30). 
 Therefore,  the VEV of  $\vec n(x)$ is given by 
 \begin{eqnarray}
\langle \vec n(x) \rangle_J \sim \frac1{L\beta} \frac{\partial}{\partial \vec J_c^\dagger} \ln Z_\lambda[J] \Big|_{\rm s.p.}
=\frac {\vec J_c}{\lambda}\Big|_{\rm s.p.}
\end{eqnarray}
("s.p." stands for the saddle-point-value) 
 which  obviously vanishes  as long as the mass gap remains nonzero in the limit $\vec J_c \to 0$.   

Next  let us reconsider the gap equation:
\begin{eqnarray}
0=\frac{\partial F_\lambda[J]}{\partial \lambda} =\frac{\partial F_\lambda}{\partial \lambda} +\frac{|\vec J_c|^2}{\lambda^2} L\;.
\end{eqnarray}
Here one can again show that 
\begin{eqnarray}
\frac{\partial^2 F_\lambda[J]}{\partial \lambda^2} <0\,,\quad   \lim_{\lambda \to 0}  \frac{\partial F_\lambda[J]}{\partial \lambda}  =\infty\,,\quad
\lim_{\lambda \to \infty}  \frac{\partial F_\lambda[J]}{\partial \lambda} =-\infty\,,
\end{eqnarray}
which leads to the uniqueness of the solution  $\lambda=\lambda_{\rm sol}$ of the 
gap equation,   even in the presence of an arbitrary $\vec J_c$.    $\lambda_{\rm sol}$ satisfies
\begin{eqnarray}
\lambda_{\rm sol} \ge   \lambda_{\rm sol} \Big|_{J=0} \ge \Lambda^2.
\end{eqnarray}
Thus 
\begin{eqnarray}
\lim_{J\to 0}  \langle \vec n(x) \rangle_J=0\;.
\end{eqnarray}
%The spontaneous symmetry breaking does not  occur. 
%Naively,  as a manner in taking Wilsonian  effective action,   
%keeping `vev's of fields as  a classical fields is considered to give the same effective action and thus the same results as the above.     
%Here we integrate out `'VEV'  $\vec \sigma$ completely.    However  the argument around the wrong vev  might gives a wrong answer.
%In Sec.3 in our paper,  we  DO partially integration,  but  do NOT  integrate a size moduli  $\hat \sigma^2$ to 
%show the reason why the work by SMY paper gives the wrong answer.
% 
%By following the same procedure as above one finds 
%\begin{eqnarray}
%\lim_{J\to 0} \langle |\vec n |^2\rangle 
% &=&\lim_{J\to 0} \frac1{Z[J]} \sum_i\frac{\delta^2}{\delta \vec J_i (x)^\dagger \delta J_i (x)}  Z[J] \nonumber \\
% &\sim &
%\lim_{J\to 0}  \frac1{Z_\lambda[J] (\beta L)^2} \sum_i  \frac{\partial^2}{\partial (J_c)_i^\dagger \partial (J_c)_i}  Z_\lambda[J] \Big|_{\rm s.p.}
%\nonumber \\
% &=&  \frac{N}{\beta L \lambda}\Big|_{\rm s.p.} 
%\end{eqnarray}
% which is exactly $\hat \sigma^2$ given in Section 3, showing the consistency.
% 
% 
%

%%%%%%%%%%%%%%%%%%%%%%%%%%%%%%%%%%%%%%%%%%%%%%%%%%%%%%%%%%%%%%%%%%%%%

\end{document}